# Relationship between Remittances and Macroeconomic Variables in Times of Political and Social Upheaval: Evidence from Tunisia's Arab Spring[1]

Jamal Bouoiyour [a] [2], Refk Selmi[a, c] and Amal Miftah[b]

[a] CATT, University of Pau, France

[b] LEDa, University of Paris-Dauphine, France

[c] University of Tunis, Tunisia

**Abstract**

If Tunisia was hailed as a success story with its high rankings on economic, educational, and other indicators compared to other Arab countries, the 2011 popular uprisings demonstrate the need for political reforms but also major economic reforms. The Arab spring highlights the fragility of its main economic pillars including the tourism and the foreign direct investment. In such turbulent times, the paper examines the economic impact of migrant' remittances, expected to have a countercyclical behavior. Our results reveal that prior to the Arab Spring, the impacts of remittances on growth and consumption seem negative and positive respectively, while they varyingly influence local investment. These three relationships held in the short-run. By considering the period surrounding the 2011 uprisings, the investment effect of remittances becomes negative and weak in the short- and medium-run, whereas positive and strong remittances' impacts on growth and consumption are found in the long term.

[1] ***Acknowledgement:*** The content of this paper has been presented at the annual seminar of ESC Pau - IRMAPE / CATT / CREG held on January 21, 2016, the 4th Transitions in Middle East and North Africa (TMENA) meeting on April 2016 in Hammamet-Tunisia and the Development seminar /GETHA LAREFI/LAM held in Bordeaux on March 4, 2016. We are so thankful for all the participants, in particular Christophe Muller, Mouhoub El Mouhoud, Antoine Bouet and Eric Rougier. The authors are also highly indebted to Diaa Noureldin, Kamiar Mohaddes, Magda Kandil, Mahmoud El-Gamal, Hoda Selim, Hany Abdel-Latif, Ghazi Ibrahim-Al Assaf and all the participants to the Macroeconomics session at the ERF 23rd Annual Conference held on March 18-20, 2017 in Amman- Jordan for helpful comments and insightful suggestions. Remaining shortcomings are the responsibility of the authors.

[2] Corresponding author: Jamal Bouoiyour; Address: Avenue du Doyen Poplawski, 64000 Pau, France; Tel: +33 (0) 5 59 40 80 01, Fax: +33 (0) 5 59 40 80 10, E-mail: jamal.bouoiyour@univ-pau.fr.

# 1. Introduction

On 17 December 2010, a young Tunisian street merchant, Mohamed Bouazizi set himself on fire that ended his life and sparked unrest sweeping in Tunisia. His tragic suicide was seen as an act of despair, humiliation and protest of the explosive problems confronted by the majority of Tunisians who were no longer prepared to accept inequalities, corruption, lack of freedoms, unemployment, etc. The winds of change that swept across Tunisia triggered a "domino" effect in different Arab countries including Egypt, Libya, Syria and Yemen. The term "Arab Spring" has come to present these popular revolutions. In the afternoon of the euphoria of the 2011 protests, Tunisia experienced an evolving volatility and growth slow-moving. Before the downfall of the 23 year-old regime of Ben Ali, Tunisia was one of the widely cited development success stories in the Middle Eastern and North African region, and was portrayed as a "top reformer" as far as institutional reform was concerned (Pollack 2011). Its economic is more prosperous with a growth rate in 2011 projected to exceed 5 percent, outpacing low-middle-income countries' averages. Thanks to the 1986 structural adjustment program and the macro-economic improvement called 'economic miracle" beginning in the late 1990s, the country has also succeeded to keep its domestic and external economic imbalances under control. Further, there have been positive advances in education and women rights. However, issues of youth unemployment, corruption, civil and political rights and unequal wealth distribution have received much less attention. In fact, despite a marked economic and educational progress, the social conditions of the Tunisian people have deteriorated, and the corruption and inequalities have reached a very high level. It comes as no surprise that popular uprising occurs in such framework.

The Arab Spring has produced immediate negative repercussions on economic development. So, there has a sharp decrease of the annual growth: 1 percent per year between 2011 and 2015. The national economic base suffered. According to the National Institute of Statistics of Tunisia, the foreign direct investment (FDI) plunged by 7.6 percent in 2016 compared to 2010. Also, tourist arrivals and revenues collapsed by 30.8 and 35.1 percent, respectively, and the dinar depreciated substantially. It is expected that a decrease in tourist arrivals can have a large effect on the Tunisian economy since tourism is a source of direct employment and foreign currency reserves. Further, the trade deficit rose markedly to reach 13.6 percent of GDP.

In order to alleviate the adverse effects of such political instability on economic outcomes, there is a need for considering counter-cyclical financing mechanisms and other pillars of Tunisia economy. The most tangible of these are migrant' remittances, the income that migrant send home, potentially leading to cushion the harmful effects of this political and social upheaval. In fact, in times of crises (2008 and 2011), remittance flows showed a resilience (World Bank 2012). Nevertheless, these financial flows had not attracted so much attention from successive Tunisian governments, unlike other countries such as Morocco where they have been and are still being one of the major sources of financing the economy (Bouoiyour 2006).

In this study, we test whether remittances may boost economic development, stabilize consumption fluctuations and stimulate investment activities with reference to the case of Tunisia witnessing the 2011 Arab Spring unrest. While a large strand of literature has focused on how remittance inflows interact with economic growth and investment (Glystos 2002, Fayissa and Nsiah 2008, Yang 2008, Tansel and Yasar 2009, Barajas et al. 2009, among others), very little was devoted in the literature to the stabilizing effects of remittances on consumption variations. In fact, One of the most threatening impacts of output shocks is the consumption instability which negatively influences agents' welfare (for instance, Bhaumik and Nugent 1999, Kedir and Girma 2003, Castaldo and Reilly 2007). Also, a limited number of studies have analyzed the ability of remittances to act as a buffer against shocks (Lueth and Ruiz-Arranz 2007, Chami et al. 2005). This paper extends previous literature in the following important aspects. First, it simultaneously examines the remittances' impacts on economic growth, domestic investment and consumption. It is necessary to note that a few attempts have been made to empirically investigate the development impacts of remittances in Tunisian case (Mesnard 2005, Jouini 2015 and Kouni 2016). Second, it seeks to identify through which channels remittances can spur Tunisia's growth during turbulent times. Third, it revisits the relationship between remittances and macroeconomic variables placing particular attention on possible nonlinear relationship. The majority of previous researches on the issue has ignored the non-linearity of the relationship between remittances and economic development or has employed a quadratic term to capture nonlinearity. With respect to the remittances effects on macroeconomy, Ruiz et al. (2009) showed a positive link between remittances and economic growth in parametric estimations, whereas such a relationship disappears when nonlinearity is taken into account using semi-parametric and non-parametric methods. Besides, Hassan et al. (2012), by analyzing the effects of inward remittances flows on per capita GDP growth in Bangladesh during the period 1974-2006, argued that the developmental impact of remittances may not be linear. Accurately, they found a U-shaped relationship that exists between remittances and long-term total factor productivity growth, where the impact of remittances flows on growth is initially negative but becomes positive later on. They attributed these outcomes to the "unproductive" use of remittances in the beginning followed by "more productive" utilization in late stages. In line with these findings, our empirical strategy seeks to verify the non-linear linkage between remittances and some macroeconomic variables. However, our approach is different to the existing literature because we are able to address such relationship in an unstable framework using a novel empirical strategy that accounts for nonlinearity pattern. To avoid misspecification biases from imposing an arbitrary functional form, we apply a new data analysis tool, namely Empirical Mode Decomposition (EMD), which decomposes each time series into a scale-on-scale basis and at each scale it is estimated the correlation. The motivation behind the use of this technique arises in the desire to extract intrinsic characteristics inherent to the time series. Prior research has been performed by employing different techniques, in particular a cointegration analysis or an autoregressive distributed lag (ARDL). Listing all existing estimators is definitely beyond the scope of this study. As the existing literature on the relationship between remittances and macroeconomic variables is rather inconclusive, it warrants for further empirical investigation. Sun and Meinl (2012) claimed that most data convey noises that are caused by intricate structure of irregularities and roughness. They thus use wavelet analysis to denoise the data and to avoid the manifold irregularities along with different time-scales and frequency components. Every component resulting from a wavelet transform has parameters that determine its scale and level over time which avoids the possible non-stationarity problem. But it would be more appropriate to have a transform that would not solely allow dealing with non-stationarity problem, but also carrying out an adaptive transform basis. A successful data assessment is heavily sensitive to the choice of data-scale representation, and its ability to provide reliable and robust data-association metrics for real data. For this purpose, it is important to account for scales which are free from rigid mathematical constraints and data driven to reflect the inherent movements embedded in the data, without a priori knowledge. EMD, in this way, has proven to be effective in a broad range of applications for extracting signals from data generated in noisy nonlinear and non-stationary processes (Huang et al. 1998, 2003; Huang and Attoh-Okine 2005). Recently, a particular attention have been given to the Empirical Mode Decomposition given its ability to disentangle any signal into its scale components, its flexibility to handle non-stationary data and its capacity to provide an alternative representation of the association structure between time series on a scale-by-scale basis.

Using a multi-scale correlation analysis via the Empirical Mode Decomposition, quite interesting results were drawn. Prior to Arab Spring, the short-term hidden factors of remittances explain negatively the economic growth, varyingly the local productive investment and positively the consumption. These results change fundamentally when accounting for the period surrounding Tunisia in the onset of Arab Spring. While the findings remain stable for the remittances-investment linkage (negative and weak and driven by short- and medium-term factors), the cycles remittances-growth and remittances-consumption became positive, greater and explained by long-term inner features. These findings are fairly robust to the control for endogeneity bias and to the use of further signal approaches.

The outline of the paper is as follows. Section 2 presents a literature review on the channels through which remittances can enhance economic growth in the developing countries. Section 3 gives some stylized facts. Section 4 discusses the methodology and provides a brief data overview. Section 5 reports and discusses our results. Section 6 checks the robustness of our findings. Section 7 concludes and offers relevant policy implications.

## 2. Literature review

In light of the increasing evidence on the substantial role of remittance flows relative to other flows in developing countries, it is not surprising that the last decade was marked by tremendous attention by policymakers and academics devoted to their developmental role. A wider macroeconomic literature has concentrated in the impact of remittances on growth, investment, consumption and monetary and exchange rate policies. The findings, nevertheless, are mixed, and sometimes controversial.

Literature has underscored various channels through which migrant remittances can spur economic growth in the developing countries. However, it has proven not easier to fully support the idea that remittances provide a boost to economic growth of recipient economies, and whether they help lighten economic hardship. Concerning this point, remittances can mitigate output growth volatility because of their relative stability. Some papers argued that remittances may act as a countercyclical stabilizer in receiving countries. For example, Chami et al. (2005) indicated that remittances have a tendency to move counter-cyclically with the GDP in recipient countries, consistently with the model's implication that remittances are compensatory transfers. However, Lueth and Ruiz-Arranz (2007) found that remittance receipts in Sri Lanka may be less shock-absorber than usually believed.

A limited strand of literature which has tested the direct relationship between remittances and economic growth typically showed "multi-sided" outcomes. Estimating panel growth regressions both on the full sample of countries (composed of 84 countries) and for emerging economies only, Barajas et al. (2009) claimed that remittances had, at best, no impact on economic growth. Fayissa and Nsiah (2010) investigated the aggregate impact of remittances on the economic growth of 18 Latin American countries for the period 1980–2005 and showed that a 10 percent increase in remittances led to a 0.15 percent increase in the GDP per capita income. Using the Solow growth model, Rao and Hassan (20012) explained the impact of remittances on growth by distinguishing between the indirect and direct growth effects. They found that these funds were likely to have a positive but modest effect on economic growth. These authors identified seven channels through which remittances could have indirect growth effects: the volatility of output growth, the exchange rate, the investment rate, the financial development, the inflation rate, the foreign direct investment and the current government expenditure. To a larger extent, the surveyed literature suggested different channels through which remittances could spur economic growth. In the short-run, remittances allowed home countries to strengthen the foreign-exchange reserves helping to adjust their economy. Nevertheless, the rather extensive literature on remittances provided further insights on the effects of remittances on consumption and investment (El-Sakka and Mcnabb 1999, Glytsos 2002). Accordingly, for a sample of five Mediterranean countries (namely Egypt, Greece, Jordan, Morocco and Portugal), Glytsos (2002) analyzed the remittances' impact on growth and deduced that the good done to growth by rising remittances is not as great as the bad done by falling remittances.

From an economic development viewpoint, a vexing question remains: how remittances are used. Are they spent on consumption, or are they used for productive investments? Remittances are generally spent on consumption but there is some evidence that in the long term international remittances may be channelled into productive investment. In this context, some studies looked into the effects of remittances on domestic investment (and hence indirectly on growth) and supported these optimistic conclusions. For example, Woodruff and Zenteno (2004) analyzed such effects using data of a survey of more than 6,000 self employed workers and small firm owners located in 44 urban areas of Mexico and estimated that more than 40 percent of the capital invested in microenterprises in urban Mexico was associated with migrants' remittances. There is also evidence supporting that return migration could increase investment in some developing countries like Egypt (McCormick and Wahba 2003, Wahba and Zenou 2009) and Tunisia (Mesnard 2004).

Potentially, in countries where access to credit is a major obstacle for entrepreneurship, return migration invigorated the propensity of returnees to become self-employed upon return but also the positive impact of accumulated savings on the decision to become self-employed. Additionally, it has been commonly argued that investment is directly linked to the development of financial system (Aggarwal et al 2006, Giuliano and Ruiz-Arranz 2009). By analyzing the remittances effects in Tunisia during the period from 1987 to 2012, Kouni (2016) argued that remittances have contributed to economic growth. The author showed that the amount of remittances allocated to investment is smaller than the remittances allocated to consumption. He also indicated that remittances played a potential role on explaining the share of the sectoral value added in GDP. In particular, a rise by about 1 percent in remittances allocated to investment increase the value added to GDP ratio by 1 percent to 4 percent.

Even though remittances allow home countries to strengthen their foreign-exchange reserves influencing their macroeconomic equilibrium and GDP growth, the rather extensive literature on remittances provides some insights about their detrimental impact on economic growth through the effect of the Dutch Disease. This could result from the reduction of the competitiveness of the tradable sector after an appreciation of the real exchange rate. This logic can be illustrated using the results reported by Amuedo-Dorantes and Pozo (2004). The authors found, for a sample of 13 Latin American and Caribbean countries, that remittances have the potential to inflict economic costs on the export sector of receiving countries by inducing a loss of international competitiveness. In the case of Tunisia, Chnaina and Makhlouf (2015) showed that an increase in worker's remittances of 1 percentage point of GDP is associated with an appreciation of Tunisia's real exchange rate by 0.39 per cent[3]. There are other channels through which remittances could affect growth, namely human capital and labor supply. Thus, remittances can stimulate investment in human capital and health as well (Mansuri 2006, Valero-Gil 2008). They may also influence economic growth via their effects on the labor force participation. However, these effects of remittances are sensitive to the considered countries. Some migration research showed a negative effect on labor supply if remittance income substitutes for labor income. They had also a disincentive impact on work and savings in the origin community of migrants i.e., the moral hazard phenomenon (Chami et al. 2005), leading to a decrease in labor supply. Nevertheless, as noted by Özden and Schiff (2006), such decline in labor supply caused by remittances may prompt high productivity.

---

[3] Bouoiyour and Selmi (2016) tested the occurrence of Dutch Disease hypothesis (i.e., whether the increase of remittance flows leads to an appreciation of the real effective exchange rate) in the Tunisian case, and provided evidence supporting such hypothesis. They further found that this effect operates strongly through the differential price and modestly via the nominal effective exchange rate.

### 3. Migration flows and remittances to Tunisia

Migrants from Tunisia are predominantly destined for Europe, and for historical and political reasons, France has attracted the majority of the Tunisian community abroad. According to the official data, 1,223,213 Tunisians (i.e. 10 percent of the Tunisian population) were residing abroad in 2012, more than 1 million of whom lived in Europe (668,668 in France). Tunisian migration flows to traditional European countries like France and Germany have increased during the last decades largely owing to family reunification, whereas migration to the other destination countries is mainly explained by labour migration. This is the case for example of the migration to Gulf countries which is generally temporary, responding to economic and political backgrounds in Tunisia and in these host countries. High unemployment and recent political instability in the country are potentially the most important reasons of emigration. Young and graduate unemployment represents a drama in the lives of many individuals in this country. The official data suggest that in 2012, graduate unemployment rates (tertiary education level), in Tunisia, stood at 26.1 percent. What is more, the high skilled emigration had grown significantly over the past two decades, reflecting the selective nature of migration by educational attainment and the general improvement in the level of education in this country. Looking at the OECD data about the emigration rate of the highly educated persons[4] in 2010-2011, Tunisia has almost 10 percent of its skilled workforce living abroad (OECD 2013). Note that during the time of the revolution, there was a significant increase of irregular migration flows towards Europe. A prominent feature linked to Tunisian migration is the important funds sent by migrants. They registered a noticeable increase during the two last decades. In 1990, international remittances received were around $0.5 billion; by 2008, this number rose to $1.9 billion. In 2014, they attained $2.35 billion. These official statistics reported by the Central Bank of Tunisia largely underestimate the total amount of migrants' remittances because Tunisian migrants frequently used informal modes of transfer. In Tunisia, informal remittances carried by travellers from Europe (migrants, family, friends and acquaintances) were estimated to account for 38 percent of total remittance receipts (IOM 2011).

---

[4] The emigration rate of highly educated persons from country i is calculated by dividing the highly educated expatriate population from country of origin i by the total highly educated native-born population. Highly educated persons correspond to those with a tertiary level of education.

The growing importance of remittances to Tunisia is reflected in Figure1 where we reported the evolution of these flows as percentage of GDP. Remittances as a share of GDP varied between 3.77 and 5.01 percent during the period 1995-2015. As such, remittance receipts might exert a significant impact on Tunisian development of Tunisia in a period of political and social upheaval. Representing one of the potential sources of foreign currency and national saving for Tunisia, these remittances inflows played a pivotal economic role in the hardship periods. In fact, remittances represented 28.7 percent of national saving in 2012.

**Figure 1. Remittances to Tunisia**

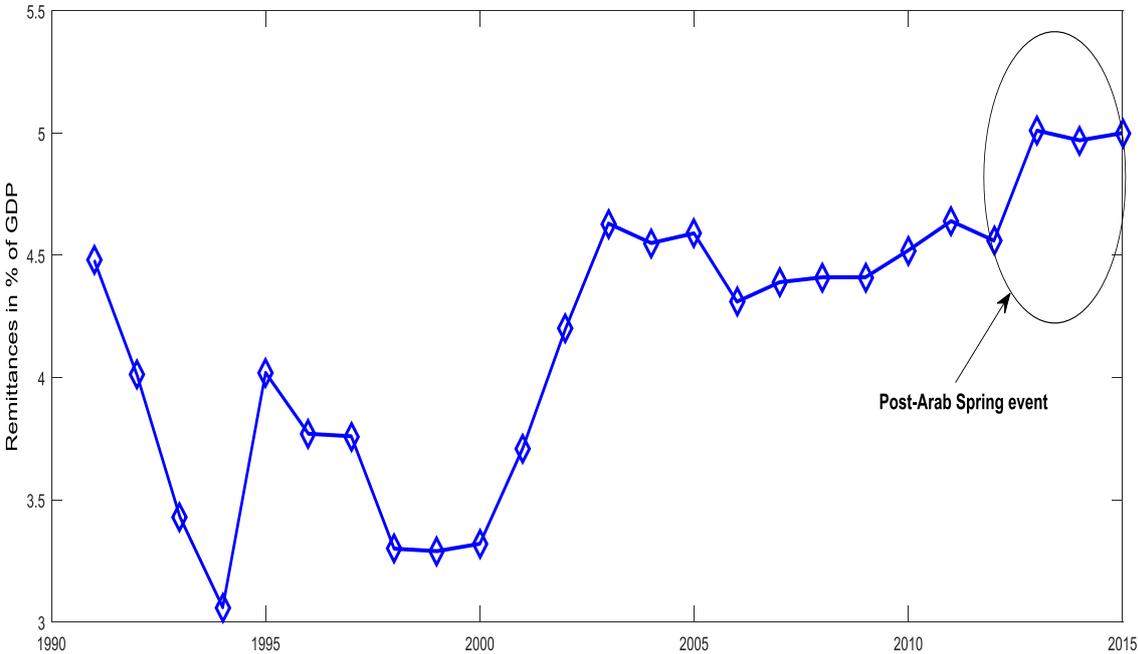

Source: World Bank.

By looking at the Figure 1, we note that neither the 2008 economic crisis nor the 2011 uprisings exerted strong influence on remittances flows from Tunisian migrants. It should be noted that remittances to Tunisia come essentially from Europe and Arab countries. In fact, they are widely originated from France, with Germany and Italy trailing far behind. Among Arab countries, the Gulf Cooperation Council countries (GCC)[5] are the main remittances sending countries followed by Libya before the Arab Spring which caused a marked decline of remittances sent from Libya.

---

[5] Within the GCC region, the main remittances sending countries in 2013 were the Saudi Arabia and the United Arab Emirates (Central Bank of Tunis 2014).

## 4. Methodology and data

It is recognized that the investigation of dynamic interactions between time series is an important issue that has long posed challenge to economic agents and academics. In investigating remittances effects, most empirical studies use techniques that look for linear positive or negative relationships. However, the relationships between remittances and macroeconomic variables may be nonlinear, especially when focusing on an unstable context. In general, the historical data of time series are the result of complex economic processes which include policy shifts, structural changes, sudden shocks, political tensions, among others. The combined influence of these various events are the root of distributional characteristics of financial and macroeconomic time series such as asymmetry, nonlinearity, heavy-tailness and extreme values. Given these considerations, the primary objective of this study is revisit the relationship between remittances, economic growth, investment, consumption and real effective exchange rate while accounting for the scale-on-scale variation (i.e., nonlinearity) and the hidden factors that may drive it.

The literature is quite rich in methods to assess time-varying correlations. The traditional time series analysis tools usually rely on Fourier transforms in one way or another. Nevertheless, according to Huang et al. (1998), the Fourier transform might prompt inaccurate information owing essentially to the nature (in the time domain) of the transform. Even wavelet analysis, developed to deal with non-stationarity and local frequency changes, produces confusing and sometimes contradicting results when applied to environment and climate signals (Sonechkin and Datsenko 2000, Oh et al. 2003). By performing wavelet approach, it is sometimes not easier to determine local frequency changes because the spectrum is generated by stepping via several predetermined frequency components showing generally blurred findings. Wavelet method has a problem of shift variance. More accurately, if the start point varies, by for example dropping the initial point, the wavelet transform may reveal distinct outcomes. However, the EMD method makes no assumption about linearity or stationarity and the intrinsic mode functions (IMFs) are often easily descibed[6]. A signal can be disentangled into a sum of finite number of zero mean oscillating components having symmetric envelopes defined by the local maxima and minima. The EMD is based on the sequential extraction of energy associated with distinct frequencies ranging from high fluctuating components (short-run) to low fluctuating modes (long-run).

---

[6] For detailed discussion of the EMD technique and comparison to other time series analysis tools, you can refer to Huang et al. (1998) and Flandrin et al. (2004).

In practice, the IMFs are extracted level by level: first, the high frequency oscillations riding on the corresponding low frequency oscillations are identified; then the next level highest-frequency local oscillations of the residual of the data are extracted. The sifting algorithm to create IMFs in EMD consists of two steps. First, the local extremes in the time series data $X(t)$ are identified. Thereafter, all the local maxima are connected by a cubic spline line $U(t)$ generating the upper envelope of the time series, and another cubic spline line $L(t)$ generating the lower envelope. For this purpose, we initially measure the mean $m_1$ for different points from upper and lower envelopes, given by:

$$m_1 = (U(t) + L(t))/2 \tag{1}$$

**Figure 2. The identification of the upper and lower envelopes and the mean**

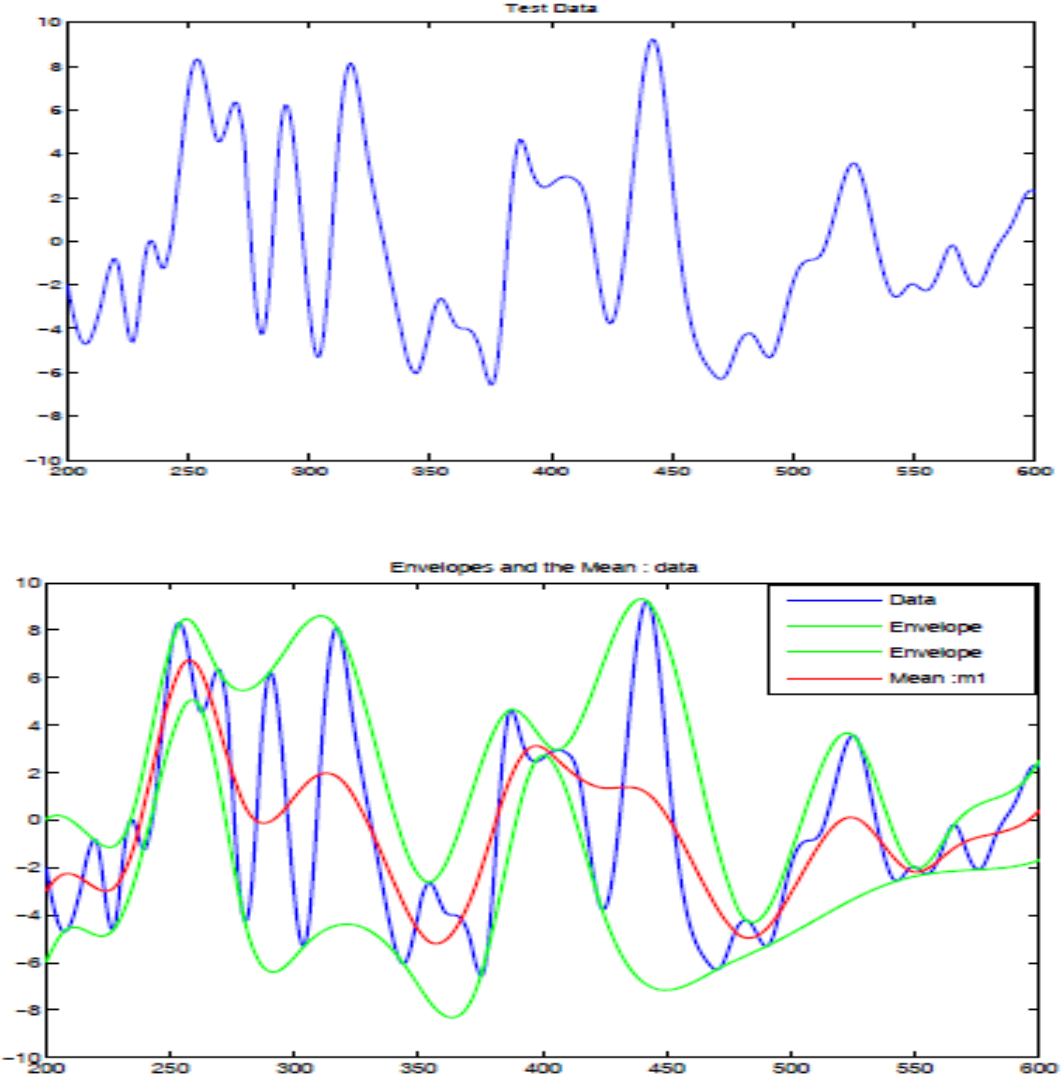

Note: The data (blue) upper and lower envelopes (green) defined by the local maxima and minima, respectively, and the mean value of the upper and lower envelopes given in red.

The difference between the original data and $m_1$ is the first component (Figure 3), called $h_1$.

$$X(t) - m_1 = h_1 \qquad (2)$$

**Figure 3. The first component: Original signal-$m_1$**

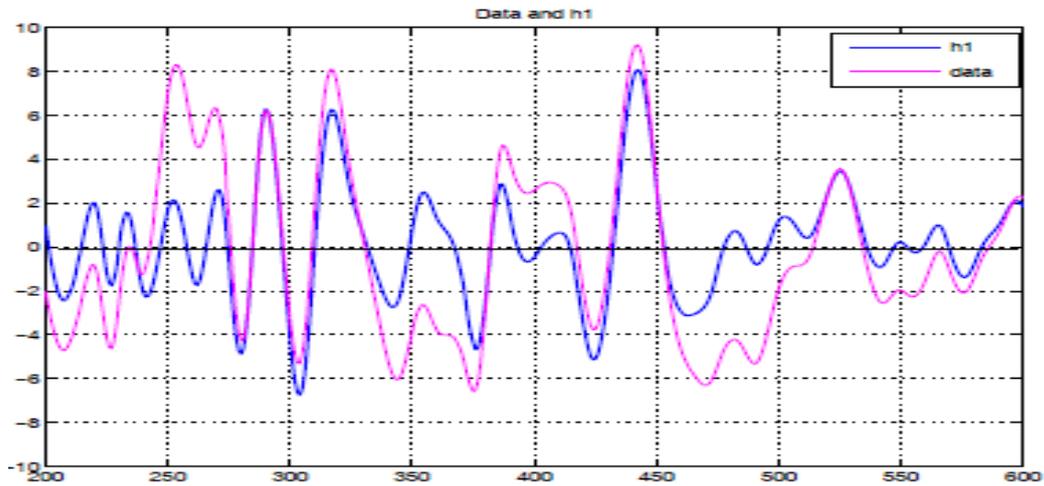

If the $h_1$ is not an IMF, we have to repeat the sifting process till it reduced to an IMF. Then, in the subsequent steps of sifting process, the first component $h_1$ is treated as if it were the data, i.e.,

$$h_1 - m_{11} = h_{11} \qquad (3)$$

**Figure 4. The sifting process**

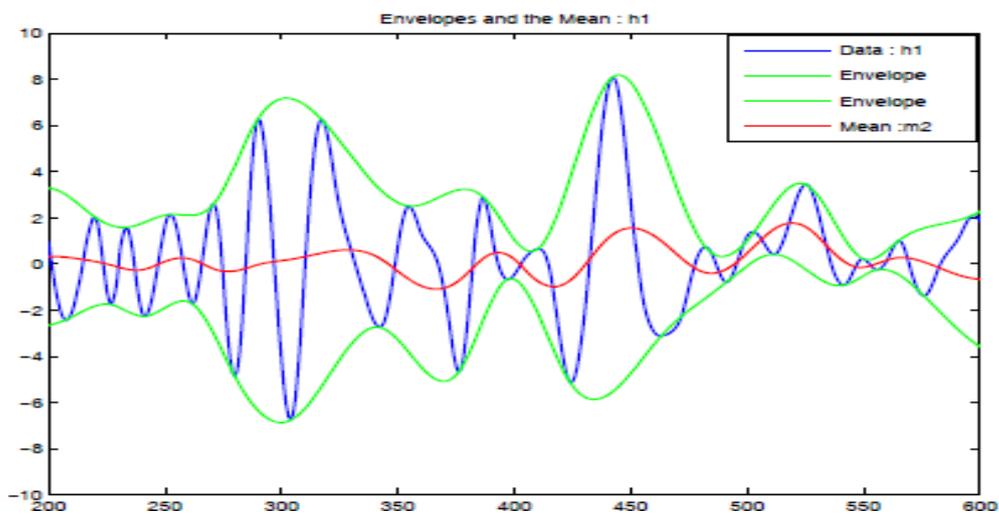

The sifting process would be done *k* times till acceptable tolerance is reached:

$h_{1(k-1)} - m_{1k} = h_{1k} = c_1$ (4)

If the resulting time series $h_{1k}$ is an IMF, then it is dubbed as $c_1$ which is the real first component which satisfies the definition of IMFs (see Figure 5). Equation (2) could be rewritten as follows:

$X(t) - c_1 = r_1$ (5)

**Figure 5. The first residual component: Original signal $-c_1$**

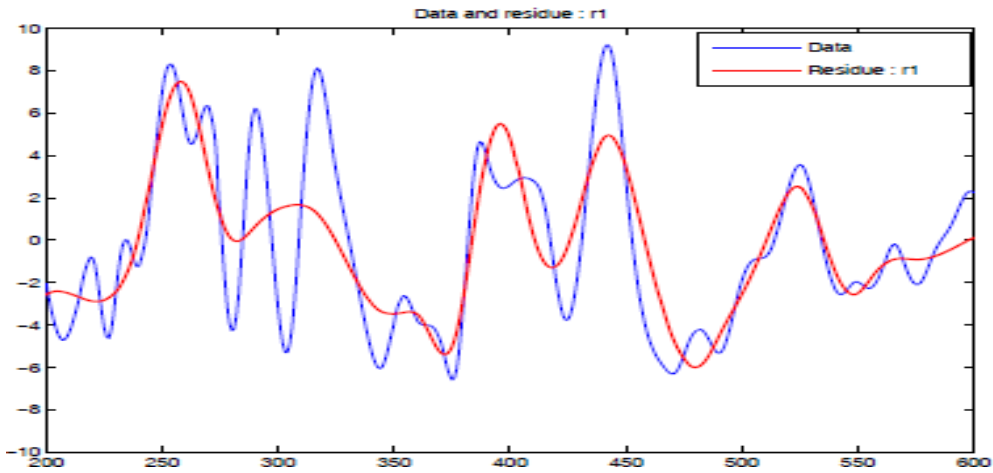

Equation (5) will also be repeated many times until the residue (*r*) becomes a monotonic function from which no more IMFs can be extracted[7]. The last residue is the trend of the data. Ultimately, equation (5) can be denoted as:

$$X(t) = \sum_{i=1}^{n} c_i + r_n$$ (6)

Summing up, the decomposition of the signal into IMFs is carried out as follows: after determining the positive peaks (maxima) and negative peaks (minima) of the original signal, we construct the lower and the upper envelopes of the signal by the cubic spline method (red). In addition, we measure the mean values (blue) by averaging the upper envelope and the lower envelope. Besides, we subtract the mean from the original signal to find the first intrinsic mode function (IMF1). Then, we calculate the first residual component by subtracting IMF1 component from the original signal. Finally, we repeat the steps above until

---

[7] For more details about the way EMD works, please refer to the following link: http://perso.ens-lyon.fr/patrick.flandrin/emd.html

the final residual component becomes a monotonic function and no more IMFs can be extracted.

After the partition of the original series into different scales each one related to different timing frames, at each scale it is estimated the correlation. By using this newly econometric tool, it is possible to denoise the original series and look at detail patterns (tees). In this way, correlation analysis-based EMD provides a rich source of potential nonlinear dynamics depicting temporal dependence. Throughout this study, we consider three regressions: (1) the regression of real per capita growth[8] (gGDP) on remittances to GDP (REM/GDP) and potential control variables commonly considered as the main determinants of economic growth, including foreign direct investments to GDP (FDI/GDP), investment to GDP (INV/GDP), credits to private sector (Credits/GDP), trade openness (or the level of exports plus imports to GDP, noted OPEN) and real effective exchange rate (REER or the ratio between prices of tradable and non-tradable goods where an increase in price of tradable goods corresponds to a real depreciation); (2) the regression of domestic investment (INV/GDP) on remittances and other explanatory variables including (FDI/GDP), gGDP, Credits/GDP, OPEN, inflation (CPI), and real interest rate (RIR); and (3) the regression of consumption to GDP (CONS/GDP) on remittances, gGDP, Credits/GDP, CPI and RIR. Because we have not enough observations to estimate after the Arab Spring, we have made two estimates for two different periods: the first one corresponds to the period before the Arab Spring spanning between 1990:Q1 and 2010:Q4 (i.e., 85 observations) and the second one refers to an extended period (prior to and post Arab Spring event) that spans between 1990:Q1 and 2015:Q3 (i.e., 104 observations).

The chosen sampling period is due to data availability. The data on remittances, investment, real per capita growth and the additional explanatory variables were collected from world development indicators (CD-ROM), quandl website and Econstats[TM]. In order to assess the dynamic dependencies (correlation and causality) among the focal variables, we have transformed the variables by taking natural logarithms to correct for heteroskedasticity and dimensional differences between the investigated time series.

---

[8] We have used population series to convert the time series into per capita.

# 5. Results

## 5.1. The decomposition of remittances and macroeconomic variables via EMD

The fundamental question of this study is beyond the classic debate which opposes the remittances impact on consumption with that on investment. This research seeks to test whether these linkages evolve over different time-scales (or frequencies). Also, it assesses to what extent does Arab Spring strength the remittance matters. Our objective is to spotlight how decomposing the variables into intrinsic mode functions can be useful in examining such relationships during turbulent times. Unlike standard methods, signal approaches (in particular, a correlation analysis-based EMD approach and a frequency domain causality test) permit to uncover the inner factors that may drive the remittances' effects on growth, investment and consumption, which would stay hidden otherwise.

Figure A.1 (Appendices) displays the EMD outcomes for the variables of interest. We show that, for the restricted and the whole period, the real per capita growth, remittances, investment and consumption were decomposed into seven IMFs plus one residue. Since the number of IMFs is limited and restricted to log2N where N is the length of data[9], the sifting processes produced only seven IMFs for each variable. All the derived IMFs were listed from high frequency component to low frequency band, and the last one is the residue. Remarkably, the frequencies and amplitudes of all the IMFs evolved over time and changed when moving from the first period (before Arab Spring) to the second period (before and after Arab Spring). As the frequency changes from high to low, the amplitudes of the IMFs become wider. We discuss three main frequency components: short-run (IMF1 and IMF2), medium-run (IMF3 and IMF4) and long-run (IMF5, IMF6 and IMF7). Table 1 presents the time scale interpretation of EMD. Since for the two considered periods, seven IMFs had been derived, the interpretation of frequency components is the same for the two investigated periods.

**Table 1. Interpretation of modes based on EMD**

| Modes | Mode-interpretation |
|---|---|
| IMF1 | Short-run: within one to two quarters |
| IMF2 | |
| IMF3 | Medium-run: above two quarters and less than three years |
| IMF4 | |
| IMF5 | Long-run: above three years |
| IMF6 | |
| IMF7 | |

---

[9] The EMD technique generates itself the modes depending to the data. For more details about data extraction, please refer to Huang et al. (2003).

Table 2 reports some measures which are given to depict more accurately the derived IMFs: the mean period of each IMF, the correlation between each IMF and the original data series and the variance percentage of each IMF. The mean period corresponds to the value obtained by dividing the total number of points by the number of peaks for each IMF. Pearson correlation and Kendall rank correlation coefficients help to determine the correlations between the various IMFs and the original data. Because IMFs are intrinsically independent, it is possible to sum up the variances and employ the percentage of variance to measure the contribution of each IMF to the total volatility of the original data set.

In doing so, we obtain quite interesting findings. Before Arab Spring, the real per capita GDP growth was highly driven by short-term inner factors (IMFs 1-2). For the whole period, the contributions of trend and long-term hidden features (IMF 6) became stronger; likewise for remittances (IMFs 1-2-3 for the restricted period and IMF7 for the prolonged period) and consumption (IMFs 1-2-3 for the period before Arab Spring and IMFs 6-7 when accounting for the aftermath of Arab Spring). Unlike, gGDP, REM/GDP and CONS/GDP, INV/GDP was likely to be sensitive to short-term factors (IMFs1-2) for the two investigated periods.

**Table 2. IMFs features**

|  | Restricted period (1990:Q1-2010:Q4) | | | | Whole period (1990:Q1-2015:Q3) | | | |
|---|---|---|---|---|---|---|---|---|
|  | Mean period | Pearson correlation | Kendall correlation | variance as % of the sum of (IMFs+residue) | Mean period | Pearson correlation | Kendall correlation | variance as % of the sum of (IMFs+residue) |
| **gGDP** | | | | | | | | |
| IMF1 | 1.33 | 0.496*** | 0.433* | 33.22% | 1.86 | 0.059 | 0.052* | 1.16% |
| IMF2 | 1.42 | 0.285* | 0.197** | 24.08% | 36.72 | 0.312*** | 0.258** | 16.17% |
| IMF3 | 4.79 | 0.104** | 0.098* | 2.51% | 8.15 | 0.132* | 0.117** | 4.76% |
| IMF4 | 6.49 | 0.169* | 0.110** | 8.03% | 5.38 | 0.099** | 0.043 | 3.81% |
| IMF5 | 9.57 | 0.095 | 0.087 | 0.98% | 2.04 | 0.062 | 0.038 | 1.78% |
| IMF6 | 13.58 | 0.088 | 0.071 | 1.57% | 39.14 | 0.456** | 0.414* | 41.56% |
| IMF7 | 18.19 | 0.103* | 0.095 | 3.87% | 12.23 | 0.113*** | 0.101* | 3.13% |
| Residue |  | 0.414** | 0.376** | 25.69% |  | 0.324** | 0.289* | 32.67% |
| **REM/GDP** | | | | | | | | |
| IMF1 | 5.00 | 0.421** | 0.309*** | 25.61% | 1.56 | 0.105*** | 0.101* | 1.849% |
| IMF2 | 8.12 | 0.376** | 0.256*** | 13.42% | 2.38 | 0.212*** | 0.196** | 18.13% |
| IMF3 | 16.77 | 0.165*** | 0.154** | 12.50% | 3.17 | 0.295*** | 0.288*** | 6.45% |
| IMF4 | 22.49 | 0.505** | 0.461* | 8.12% | 4.95 | 0.183** | 0.172* | 8.95% |
| IMF5 | 23.86 | 0.484* | 0.083 | 3.11% | 5.78 | 0.109* | 0.100** | 2.732% |
| IMF6 | 24.74 | 0.075* | 0.052 | 2.27% | 7.45 | 0.108** | 0.097** | 1.611% |
| IMF7 | 26.63 | 0.132** | 0.096*** | 1.91% | 11.69 | 0.404** | 0.387** | 24.87% |
| Residue |  | 0.410*** | 0.393*** | 13.03% |  | 0.269** | 0.261*** | 22.34% |
| **INV/GDP** | | | | | | | | |
| IMF1 | 1.87 | 0.492*** | 0.445** | 32.00% | 18.79 | 0.322*** | 0.300** | 3.13% |
| IMF2 | 7.46 | 0.397* | 0.361** | 28.73% | 20.24 | 0.292* | 0.175** | 38.67% |
| IMF3 | 8.53 | 0.158* | 0.143** | 15.89% | 26.12 | 0.101** | 0.069** | 39.19% |

| | | | | | | | |
|---|---|---|---|---|---|---|---|
| IMF4 | 10.29 | 0.098 | 0.065 | 1.43% | 9.08 | 0.123*** | 0.119** | 1.52% |
| IMF5 | 11.37 | 0.124* | 0.115** | 1.15% | 13.72 | 0.162** | 0.135*** | 1.93% |
| IMF6 | 16.85 | 0.092* | 0.088* | 1.26% | 6.56 | 0.114* | 0.097* | 0.95% |
| IMF7 | 24.56 | 0.054 | 0.039 | 2.84% | 14.15 | 0.095* | 0.076** | 1.05% |
| Residue | | 0.102 | 0.084 | 3.09% | | 0.303* | 0.281* | 18.51% |
| **CONS/GDP** | | | | | | | | |
| IMF1 | 3.29 | 0.333** | 0.328*** | 35.16% | 4.21 | 0.112 | 0.099 | 0.68% |
| IMF2 | 5.88 | 0.197** | 0.169* | 18.42% | 5.16 | 0.109** | 0.076 | 0.56% |
| IMF3 | 8.17 | 0.168*** | 0.154*** | 16.12% | 6.10 | 0.086 | 0.054 | 0.12% |
| IMF4 | 10.46 | 0.117* | 0.103 | 6.05% | 8.93 | 0.131** | 0.116** | 7.14% |
| IMF5 | 10.93 | 0.068 | 0.045 | 0.78% | 15.34 | 0.195*** | 0.167*** | 9.23% |
| IMF6 | 11.76 | 0.104 | 0.092* | 2.08% | 16.47 | 0.262*** | 0.199** | 30.97% |
| IMF7 | 12.54 | 0.044 | 0.036 | 0.28% | 19.58 | 0.203*** | 0.197* | 22.03% |
| Residue | | 0.172** | 0.168*** | 16.68% | | 0.256* | 0.234** | 28.15% |

Notes: ***, ** and *: Correlations are significant at the levels of 1%, 5% and 10%, respectively (2-tailed).

The findings reported in Table 3 give more precise information about the three main mono-components (short- and long-term factors) determining growth, remittances, investment, consumption and real effective exchange rate, and sustain the aforementioned outcomes displayed in Table 2. We find that the contributors of the variation of the variables of interest change by moving from the restricted to the whole period, with the exception of INV/GDP. The latter is still driven by high fluctuating components during the two investigated periods. For the rest of variables, the quickly fluctuating oscillations seem the major driving factors in the restricted period, while the long-term factors determine their variations when considering the onset of Arab Spring (i.e., the whole period). However, the investment to GDP appears driven by high frequency component for the two periods.

**Table 3. Correlations and variance of components**

| | Restricted period (1990:Q1-2010:Q4) | | | Whole period (1990:Q1-2015:Q3) | | |
|---|---|---|---|---|---|---|
| | Pearson correlation | Kendall correlation | variance as % of the sum of WDFs | Pearson correlation | Kendall correlation | variance as % of the sum of WDFs |
| **gGDP** | | | | | | |
| High frequency component | 0.325* | 0.318** | 57.96% | 0.113*** | 0.077 | 6.89% |
| Low Frequency component | 0.279*** | 0.256*** | 5.72% | 0.398** | 0.367** | 59.11% |
| Trend component | 0.108* | 0.102** | 25.69% | 0.313** | 0.300* | 32.67% |
| **REM/GDP** | | | | | | |
| High frequency component | 0.412* | 0.373** | 45.62% | 0.217** | 0.181*** | 11.08% |
| Low Frequency component | 0.169* | 0.123* | 12.14% | 0.455*** | 0.424** | 49.92% |
| Trend component | 0.357*** | 0.329*** | 23.03% | 0.398** | 0.372*** | 22.34% |
| **INV/GDP** | | | | | | |
| High frequency component | 0.467** | 0.389* | 51.23% | 0.523** | 0.510*** | 63.04% |
| Low Frequency component | 0.081 | 0.064 | 8.45% | 0.131* | 0.092 | 4.12% |
| Trend component | 0.329*** | 0.296** | 13.09% | 0.267* | 0.195* | 18.51% |
| **CONS/GDP** | | | | | | |
| High frequency component | 0.481** | 0.295* | 48.78% | 0.123** | 0.110*** | 10.98 |
| Low Frequency component | 0.116* | 0.100* | 11.21% | 0.411*** | 0.372*** | 46.72% |
| Trend component | 0.398*** | 0.354** | 16.68% | 0.267* | 0.195* | 28.15% |

Notes: ***, ** and *: Correlations are significant at the levels of 1%, 5% and 10%, respectively (2-tailed).

Figure 6 indicates that each component explaining the evolution of gGDP, REM/GDP, INV/GDP and CONS/GDP exhibits dissimilar characteristics. Consistently with the aforementioned outcomes, for the restricted period (left side graph) economic growth, remittances and consumption appeared driven by high frequency components, while they seemed determined by low frequency component over the whole period (right side graph). Nevertheless, for the two periods under study, investment was determined by short-run features.

Despite the meaningfulness of the above results, it is important to determine whether there exist hidden factors driving the relationships between remittances and macroeconomic variables (i.e., growth, investment and consumption) rather than identifying what drive the time series separately. So, our main purposes are (1) to assess whether the relationship between remittances and these macroeconomic variables is time-varying, and (2) to examine whether the effect of remittances on Tunisia's growth may differ from the period prior to Arab Spring to the period post-2011 uprisings, and from one scale to another. To this end, we use a scale-on-scale correlation analysis while addressing the endogeneity problem.

**Figure 6. The hidden characteristics of the variables of interest**

| Restricted period (1990:Q1-2010:Q4) | Whole period (1990:Q1-2015:Q3) |

### gGDP

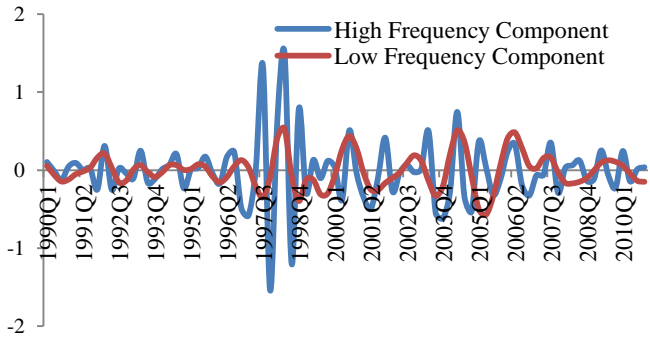 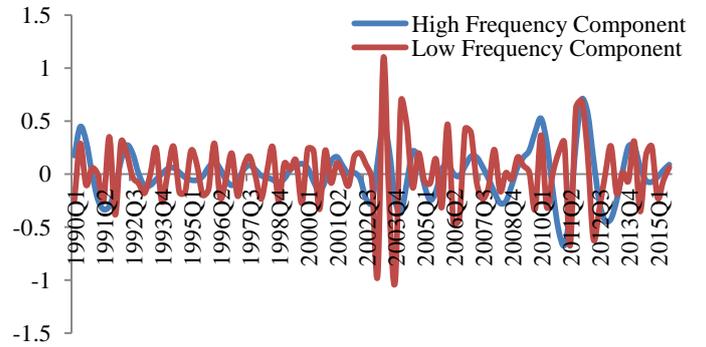

### REM/GDP

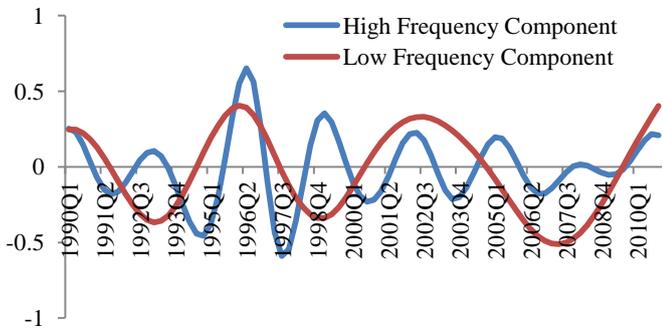 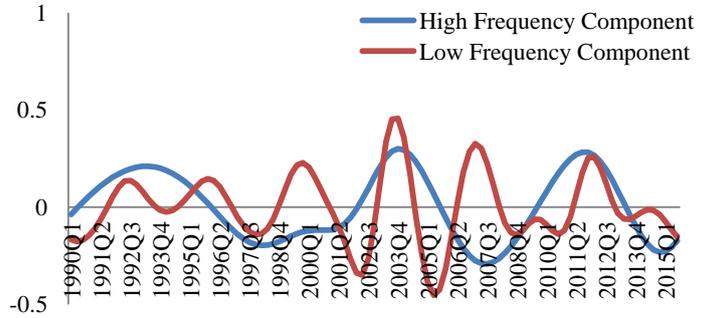

### INV/GDP

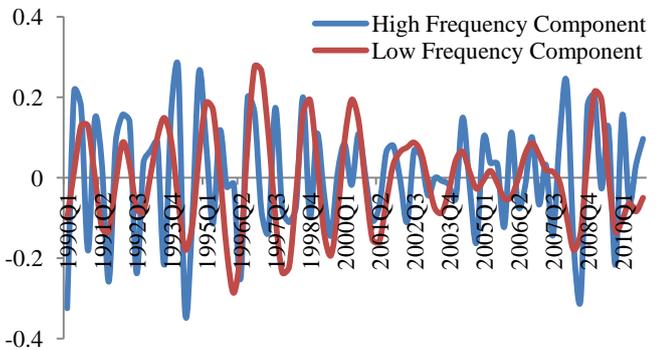 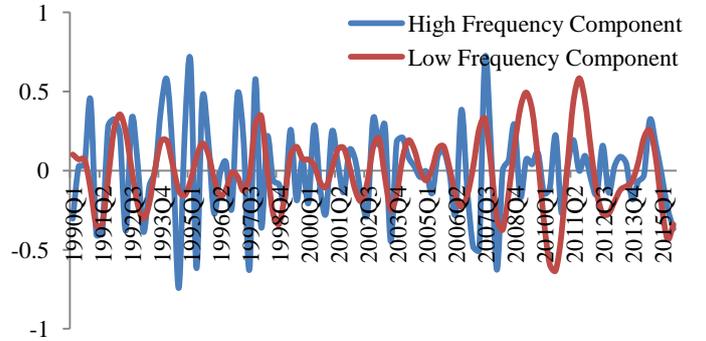

### CONS/GDP

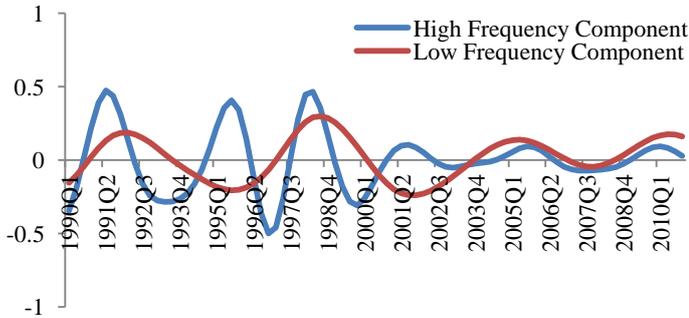 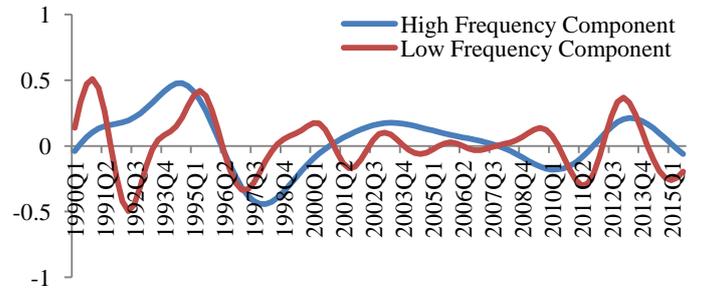

## 5.2. A correlation analysis-based EMD

We use an OLS-based EMD to assess the dynamic dependencies among remittances flows and macroeconomic variables in unstable context. Our procedure consists of regressing remittances on gGDP, INV/GDP and CONS/GDPeven if we account for potential control variables in both time domain (i.e., whole period) and among different time scales. This exercise aims at having a case of benchmarking to compare the time domain analysis with the multi-scale investigation.

### *5.2.1. Remittances and growth*

Table 4 summarizes the estimates related to the relationship between remittances and economic growth for time domain and across different time scales. Based on the time-domain analysis, we note that the remittances have no significant influence on economic growth over the restricted period (i.e. before the onset of the Arab Spring), while the effect appears significant and weaker when accounting for the post Arab Spring period (i.e., whole period). Dissimilar results are found when conducting a multi-scale analysis, highlighting that the relationship between remittance flows and growth is time-varying. In particular, the relationship is negative, weak and occurred in the medium-run (IMFs 3-4) during the restricted period. However, for the whole period (before and after the Arab Spring), remittances exert a positive and significant impact on Tunisia's growth; such relationship is dominantly driven by long-term hidden factors (IMFs 5-7). It is true that remittance flows have never been considered as a strategic variable in the Tunisian economic policy. When comparing Tunisia to Morocco, the strategic path towards migration and remittances seems totally opposed. Unlike Tunisia, Morocco has conducted an "aggressive" policy aimed at attracting remittances via the establishment of organizations dedicated to migration (such as Ministry in Charge of Moroccans living abroad, Council for the Moroccan Communities abroad, etc.). It is important to mention that the economic situation of both countries is radically different. Before the Arab Spring, Tunisia witnessed a stable economic and political conditions and strong growth. Foreign investors tended to settle easily. The openness policy has played a vital role in boosting the development of a solid and innovative manufacturing industry. This is why, Tunisia was the "champion" compared to the rest of the MENA region and a "good student" according to World Bank and IMF criteria. However, this opulence masked the existing reality of corruption and inequalities that have played a great role in the popular uprising which actually aggravated the socioeconomic situation that motivated it.

Morocco, for its part, was characterized by a stable political situation, a great resilience in dealing with external shocks (2008 economic crisis and Arab Spring in particular), but its growth is volatile due to its rain-fed agriculture. Our results support that remittance inflows to Tunisia can be served as a countercyclical stabilizer and a shock-absorber. They show that before the Arab Spring, remittances had a negative and medium-run (IMF4) influence on growth, whereas its effect in the whole period was positive and determined by long-term factors (IMFs5-7).

**Table 4. Regression of economic growth on remittances**

|  | Time domain | IMF1 | IMF2 | IMF3 | IMF4 | IMF5 | IMF6 | IMF7 |
|---|---|---|---|---|---|---|---|---|
| | **Restricted period (1990:Q1-2010:Q4)** | | | | | | | |
| C | 4.5521** | 5.328** | 5.134*** | 5.179*** | 5.092** | 5.137*** | 5.634*** | 5.553*** |
| | (2.689) | (2.976) | (4.268) | (3.768) | (2.915) | (4.118) | (4.348) | (4.492) |
| REM/GDP | 0.1157 | 0.0389 | 0.1234 | **-0.0161*** | **-0.0258*** | **-0.0251**** | 0.0682 | 0.0689 |
| | (1.575) | (1.542) | (1.376) | **(-1.863)** | **(-1.779)** | **(-2.359)** | (1.158) | (1.109) |
| FDI/GDP | 0.0718* | 0.0629** | 0.0914* | -0.0124 | 0.0393* | 0.0697** | 0.0332* | 0.0617* |
| | (1.862) | (2.698) | (1.976) | (-1.356) | (1.791) | (2.638) | (1.719) | (1.935) |
| INV/GDP | 0.0389** | 0.1255 | 0.1345 | 0.0562** | 0.1002* | 0.0411** | 0.0876* | 0.1157** |
| | (2.671) | (1.469) | (1.387) | (2.943) | (0.079) | (3.017) | (1.923) | (2.814) |
| OPEN | 0.1145*** | 0.0651 | 0.145*** | 0.0134 | -0.0188 | 0.0098** | 0.0410 | 0.0367 |
| | (3.815) | (0.589) | (3.542) | (1.156) | (0.706) | (2.923) | (1.067) | (1.156) |
| Credits/GDP | 0.0924* | 0.1155* | 0.1094** | 0.1561 | -0.1345 | -0.2671 | -0.532 | 0.2619 |
| | (1.723) | (1.914) | (2.619) | (0.956) | (-0.546) | (-1.423) | (-1.493) | (1.433) |
| REER | -0.1568** | -0.098** | -0.0862** | -0.193*** | -0.1724* | -0.146*** | -0.208** | -0.2095** |
| | (-2.492) | (-2.517) | (-2.678) | (-4.562) | (-1.854) | (-3.617) | (-2.775) | (-2.813) |
| R2 | 0.85 | 0.84 | 0.88 | 0.86 | 0.88 | 0.89 | 0.86 | 0.88 |
| | **Whole period (1990:Q1-2015:Q3)** | | | | | | | |
| C | 3.892*** | 4.892*** | 4.689*** | 4.159** | 4.356** | 5.102*** | 4.814*** | 4.415*** |
| | (3.759) | (5.168) | (4.689) | (3.029) | (3.145) | (3.924) | (4.189) | (3.624) |
| REM/GDP | 0.0054* | 0.0135 | 0.0324 | 0.0145 | 0.0357 | **0.0452*** | **0.0877**** | **0.1095*** |
| | (3.589) | (1.673) | (1.649) | (1.427) | (1.126) | **(2.789)** | **(3.524)** | **(1.823)** |
| FDI/GDP | 0.0913* | 0.0675** | 0.0532 | 0.0372 | 0.045 | 0.0479 | 0.0572** | 0.0652** |
| | (2.014) | (2.435) | (1.432) | (1.542) | (1.601) | (1.134) | (2.517) | (2.617) |
| INV/GDP | 0.0135** | 0.0102 | 0.0562 | 0.0113 | 0.055** | 0.0276 | 0.0478** | 0.0697* |
| | (2.518) | (1.459) | (1.398) | (1.3185) | (2.567) | (1.792) | (2.610) | (1.886) |
| OPEN | 0.1052*** | 0.068*** | 0.0912** | 0.0625*** | 0.084*** | 0.136** | 0.1345*** | 0.0965** |
| | (3.710) | (4.563) | (2.651) | (4.298) | (3.498) | (2.594) | (4.126) | (2.345) |
| Credits/GDP | 0.0641* | 0.0723* | 0.0542** | 0.0651* | 0.0489* | 0.0469** | 0.0345* | 0.0452** |
| | (1.865) | (1.875) | (2.921) | (1.932) | (1.932) | (2.765) | (1.699) | (2.610) |
| REER | -0.134*** | -0.197** | -0.267** | -0.2452*** | -0.189* | -0.072 | -0.078** | -0.065** |
| | (-3.772) | (-2.514) | (-2.498) | (-4.092) | (-1.796) | (-1.605) | (-2.501) | (-2.708) |
| R2 | 0.86 | 0.88 | 0.85 | 0.87 | 0.84 | 0.87 | 0.88 | 0.84 |

Notes: ***, ** and * imply significance at the 1%, 5% and 10%, respectively.

Further, during the period prior to the Arab Spring, the FDI had a positive and significant impact on growth among different time horizons. Foreign investors were highly attracted by the political stability and the high growth. However, during the uncertainty surrounding Tunisia in the onset of Arab Spring, FDI's impact on gGDP became very volatile; it was likely to be negative and positive depending to IMFs variation, but what appears meaningful is that the FDI effects fell considerably by moving from the restricted to the whole period. This outcome may be explained by the deterioration of Tunisian security situation and the lack of medium and long-term economic visibility.

Our previous results indicate that remittances help to promote economic growth during turbulent times. It remains to address whether remittances are spent on consumption, or channelled into productive investment. To this purpose, we regress investment and consumption on remittances and other relevant control variables.

### 5.2.2. *The uses of remittances: productive investment vs. consumption*

A further step consists on analyzing the relationship between (1) remittances and domestic investment to GDP, and (2) remittance inflows and consumption to GDP. The time domain and scale-on-scale results of the regression of investment on remittances are summarized in Table 5.

**Table 5. Regression of investment on remittances**

|  | Time domain | IMF1 | IMF2 | IMF3 | IMF4 | IMF5 | IMF6 | IMF7 |
|---|---|---|---|---|---|---|---|---|
|  | Restricted period (1990:Q1-2010:Q4) | | | | | | | |
| C | 2.4561** (2.651) | 1.9203 (1.122) | 1.3803 (1.327) | 1.8219 (1.266) | 2.0042 (1.523) | 3.655*** (3.254) | 4.325*** (3.645) | 1.8023** (2.895) |
| REM/GDP | -0.0345* (1.692) | **-0.0763*** **(-1.812)** | **-0.0807*** **(-1.942)** | **-0.0621*** **(-1.734)** | 0.0187 (0.121) | 0.0210 (0.112) | 0.0200 (1.161) | 0.5723 (0.408) |
| FDI/GDP | -0.0167** (-2.501) | 0.0353 (0.597) | -0.0550* (-1.841) | -0.0759* (-1.871) | 0.0138 (0.440) | 0.0833 (0.654) | -0.016*** (-3.176) | -0.027** (-2.358) |
| gGDP | 0.0245*** (3.659) | 0.0134 (0.703) | -0.0106 (0.801) | -0.0073 (0.870) | 0.0093 (0.553) | -0.0201 (0.736) | 0.0183 (1.297) | 0.0363* (1.761) |
| OPEN | 0.06239* (1.876) | 0.0932 (1.213) | 0.0763** (2.451) | 0.0764* (1.893) | 0.4321 (1.279) | 0.0679* (1.843) | 0.1389 (1.267) | 0.1056** (2.418) |
| Credits/GDP | 0.0196* (1.838) | 0.3167 (1.512) | 0.1982 (1.367) | 0.0113* (1.768) | 0.0345** (2.456) | 0.0452** (2.138) | 0.0512* (1.913) | 0.1567 (1.083) |
| CPI | -0.093*** (-3.404) | -0.1698** (-2.595) | -0.1690** (-2.552) | -0.1777** (-2.689) | -0.1118* (-1.729) | 0.1393* (1.912) | 0.0048 (0.873) | -0.0194 (0.512) |
| RIR | -0.1934** (-2.671) | -0.211*** (-4.231) | -0.222*** (-3.761) | -0.220*** (-3.6251) | -0.217*** (-4.118) | -0.195*** (-3.672) | -0.061*** (-4.110) | -0.05*** (-3.819) |
| R2 | 0.79 | 0.89 | 0.87 | 0.84 | 0.80 | 0.75 | 0.92 | 0.95 |
|  | Whole period (1990:Q1-2015:Q3) | | | | | | | |
| C | 6.8729** (2.597) | 7.5233*** (3.562) | 7.6826** (2.675) | 8.3058* (1.672) | 8.6777*** (3.845) | 8.6513*** (3.345) | 1.5678 (1.004) | 7.5233* (1.976) |
| REM/GDP | 0.0862 (1.542) | -0.452 (-1.328) | **-0.123*** **(-2.514)** | 0.2816 (0.252) | -0.1377 (-0.839) | 0.0184 (1.037) | -0.0070 (-0.982) | 0.2815 (0.276) |
| FDI/GDP | -0.0324* | 0.0165 | -0.0321* | 0.0432* | -0.0020 | -0.0125 | -0.0106 | -0.0165* |
| | (-1.810) | (0.015) | (-1.834) | (-1.697) | (-0.730) | (-0.170) | (-0.318) | (-2.132) |
| gGDP | 0.0453* | 0.0421 | -0.0120 | 0.0096 | 0.0074 | 0.0881** | 0.0686*** | 0.1345 |
| | (1.769) | (0.275) | (0.192) | (0.184) | (0.372) | (2.545) | (2.632) | (1.307) |
| OPEN | 0.1042** | 0.1084* | 0.0452*** | 0.0333*** | 0.0371*** | 0.1097* | 0.0817* | 0.1084 |
| | (2.610) | (1.884) | (3.551) | (4.162) | (3.742) | (1.941) | (1.876) | (1.221) |
| Credits/GDP | 0.0432** | 0.0568* | 0.4135 | 0.0755* | -0.0658 | -0.0612 | 0.0157** | 0.1414 |
| | (2.619) | (1.899) | (0.522) | (2.066) | (-0.920) | (-0.931) | (3.008) | (0.752) |
| CPI | -0.0368** | -0.0216* | 0.0258 | -0.0130 | -0.0030 | -0.0251* | 0.0194 | -0.0216 |
| | (-2.491) | (-2.093) | (0.273) | (-0.528) | (0.898) | (-1.876) | (0.532) | (-1.133) |
| RIR | -0.032*** | -0.0121* | -0.0370 | -0.0183* | -0.0023 | -0.0070 | -0.1223* | -0.0121* |
| | (-3.425) | (-1.698) | (-0.213) | (-2.083) | (-0.934) | (-0.807) | (-1.765) | (-1.945) |
| R2 | 0.91 | 0.95 | 0.94 | 0.96 | 0.95 | 0.91 | 0.85 | 0.95 |

Notes: ***, ** and * imply significance at the 1%, 5% and 10%, respectively.

From the time domain analysis, we note that remittances exert a negative influence on investment prior to the Arab Spring, and insignificant effect when accounting for the period after 2011 uprisings. From the multi-scale investigation, different outcomes were gathered: For the two periods under study, the linkage between REM/GDP and INV/GDP seemed to be driven by short-term factors (IMFs1-3). In terms of the sign of the remittances' coefficient, we note some changes by moving from the restricted to the lengthy period. Before the Arab Spring, the remittances effect on investment to GDP was variant (negative for IMF2 and IMF3, and positive for IMF1), while its influence was statistically negative and significant (IMF2) when considering the whole period (prior to and post Arab Spring). These findings suggest that Tunisians living abroad send their money to support their families and not for investment opportunities. These findings also underscore the usefulness of correlation analysis-based EMD when assessing remittances-investment nexus.

Table 6 reports the time domain and the multi-scale correlation outcomes of the regression of consumption on remittance inflows. All the findings go in the same direction that remittances have a positive impact on consumption either for the restricted or the prolonged period; But the correlation results derived from EMD appear more fine as we can see when exactly the relation in question is positive and when it is insignificant. Specifically, a positive link between the focal variables was found in short-run (IMFs1-2) over the period before the aftermath of Arab Spring. However, by considering the post Arab Spring period, we show that the impact of remittances on consumption became positive and more pronounced (i.e., driven by long-term inner features: IMFs6-7). Potentially, a sharp complementarity among the cycles remittances-growth and remittances-consumption was shown, sustaining the evidence remittances to Tunisia had mostly been spent for excessive consumption rather than the improvement of national investment.

**Table 6. Regression of consumption on remittances**

|  | Time domain | IMF1 | IMF2 | IMF3 | IMF4 | IMF5 | IMF6 | IMF7 |
|---|---|---|---|---|---|---|---|---|
|  | | Restricted period (1990:Q1-2010:Q4) | | | | | | |
| C | 4.521** | 4.458*** | 4.430*** | 4.422*** | 4.563*** | 4.5801** | 4.545** | 4.467*** |
|  | (2.814) | (3.456) | (4.115) | (3.629) | (3.515) | (2.764) | (1.986) | (3.197) |
| REM/GDP | 0.043* | **0.086*** | **0.089*** | 0.0035 | 0.0130 | -0.0044 | 0.022 | 0.024 |
|  | (1.892) | **(1.823)** | **(1.802)** | (0.661) | (0.602) | (0.772) | (0.697) | (0.661) |
| Credits/GDP | 0.134* | 0.168*** | 0.1525** | 0.1475* | 0.0972* | 0.0450 | -0.049 | 0.046 |
|  | (1.715) | (3.245) | (2.671) | (1.796) | (2.043) | (0.436) | (-0.368) | (0.228) |
| gGDP | 0.031** | 0.020* | 0.0345 | 0.0353 | 0.0174* | 0.0286 | 0.052 | 0.067 |
|  | (2.671) | (1.979) | (0.448) | (0.445) | (2.101) | (0.607) | (1.267) | (0.226) |
| CPI | -0.196** | -0.27*** | -0.261*** | -0.25*** | -0.243** | -0.21*** | -0.1*** | -0.213** |
|  | (-2.871) | (-3.149) | (-4.005) | (-3.814) | (-2.976) | (-4.116) | (-3.812) | (-2.689) |
| RIR | -0.062* | -0.045 | -0.0370** | -0.0329* | -0.0121 | -0.0068 | -0.005 | -0.005* |
|  | (-1.967) | (-1.531) | (-2.678) | (-1.985) | (-0.459) | (-0.691) | (-0.710) | (-2.038) |
| R2 | 0.86 | 0.90 | 0.86 | 0.84 | 0.83 | 0.87 | 0.83 | 0.79 |
|  | | Whole period (1990:Q1-2015:Q3) | | | | | | |
| C | 4.169*** | 4.469*** | 3.907*** | 3.979*** | 1.616 | 11.83** | 1.429 | 6.283*** |
|  | (3.841) | (4.576) | (3.763) | (3.986) | (1.156) | (2.561) | (1.514) | (3.612) |
| REM/GDP | 0.050** | 0.034 | 0.0310 | 0.0806 | 0.0226 | 0.0635 | **0.097*** | **0.115*** |
|  | (2.687) | (0.321) | (1.001) | (0.399) | (0.273) | (0.340) | **(1.876)** | **(2.834)** |
| Credits/GDP | 0.095 | 0.122* | 0.1573 | 0.1095* | 0.0339 | 0.0211 | 0.007 | 0.334 |
|  | (1.115) | (1.916) | (0.004) | (1.928) | (0.162) | (0.274) | (0.653) | (0.515) |
| gGDP | 0.041* | 0.022** | 0.0504* | 0.8952 | 0.3240 | 0.0261 | 0.018 | 0.122 |
|  | (1.705) | (2.397) | (1.886) | (0.450) | (0.210) | (0.117) | (0.228) | (0.786) |
| CPI | -0.076** | -0.200 | -0.109*** | 0.5006 | -0.0239 | -0.0135 | -0.004 | -0.183** |
|  | (-2.631) | (-0.963) | (-3.658) | (0.260) | (0.425) | (0.570) | (0.840) | (-2.356) |
| RIR | -0.071** | -0.056* | -0.068*** | -0.0568 | -0.08*** | -0.072** | -0.105* | 0.141 |
|  | (-1.642) | (-1.765) | (-3.914) | (-1.119) | (-4.112) | (-2.334) | (-1.921) | (-1.196) |
| R2 | 0.89 | 0.91 | 0.95 | 0.99 | 0.99 | 0.86 | 0.92 | 0.90 |

Notes: ***, ** and * imply significance at the 1%, 5% and 10%, respectively.

## 6. Robustness

There exist different ways to ascertain whether our results are fairly solid. Throughout the rest of our study, we specify two set of robustness check. First, we control for possible endogeneity bias via 2SLS-based EMD. Second, for the majority of studies on the relationship between remittances and economic development, the main question to be answered is framed around whether remittances are a statistically significant factor in boosting economic development. Another quite interesting question in relation to remittances and economic development should be that of causation. Such a question asks whether remittance flows cause economic development or visa-versa. So, because correlation does not imply causation, another focus of this study is to verify whether there exists a cyclical causal relation between remittances and the focal macroeconomic variables (growth per capita, investment to GDP and consumption to GDP). For this purpose, we utilize a frequency domain causality test[10]. The frequency domain analysis offers an appropriate alternative tool by examining the causality in frequency domain, while standard causality tests focus only on the time domain.

### 6.1. Endogeneity

The endogeneity bias is one of the methodological challenges that confront research on international migration and remittances. This can occur if remittances are sent to home country for altruistic motives or if there is an increase in workers' remittances coincided with a rise in migration from countries with low economic growth. A way to correct for the endogeneity biases is to carry out two-stage least squares (2SLS) or GMM using lag of the explanatory variables as instruments (see, for example, Giuliano and Ruiz-Arranz 2009 and Barajas et al. 2009). In the current study, we apply a 2SLS-based EMD to re-analyze the dynamic dependency between remittances inflows and macroeconomic variables in an unstable context, while controlling for endogeneity problem. We summarize the 2SLS-based EMD findings of the regressions of growth, investment and consumption on remittances and further explanatory variables in Tables 7, 8 and 9 respectively. Our results robustly reveal that before the onset of Arab Spring, remittances affected negatively the per capita economic growth and positively the consumption; such relationships held in the short- or the medium-run. However, we note a time-varying impact of these financial flows on domestic investment; it was negative in some IMFs (IMF2) and positive in others (IMF1), but it was likely to be significant only in the short-term. By accounting for the post-Arab Spring period, the investment effect of remittance inflows became weaker and determined by short- and medium-term factors, while a positive, strong and long-run remittances' effects on growth and consumption were found. Moreover, our findings also unambiguously show that either considering the restricted or the whole period, an increase in remittances is significantly linked to an appreciation of real effective exchange rate; such relationship is validated at longer time horizons. These outcomes seem consistent with the findings derived from the

---

[10] While EMD is performed within a discrete time framework, the frequency domain causality has a spectral content across a continuous range. The frequency domain causality test provides clearer cycle information almost in real time, while business cycles cannot be identified before a cycle has been completed.

OLS-based EMD, and confirm the effectiveness of the scale-on-scale correlation analysis compared to the time domain assessment[11] (Tables 4, 5, 6 and 7).

**Table 8. Regression of economic growth on remittances (Control for endogeneity)**

|  | Time domain | IMF1 | IMF2 | IMF3 | IMF4 | IMF5 | IMF6 | IMF7 |
|---|---|---|---|---|---|---|---|---|
| | **Restricted period (1990:Q1-2010:Q4)** | | | | | | | |
| C | -3.6942* | -4.894 | -6.488 | -8.207 | -6.752 | -3.553 | -2.086 | -5.346 |
|  | (-1.699) | (-1.324) | (-1.474) | (-1.532) | (-1.347) | (-0.753) | (-0.493) | (-1.236) |
| REM/GDP | -0.0034* | -0.020 | -0.012 | 0.007 | **-0.019\*\*** | -0.014 | 0.067 | -0.009 |
|  | (-1.782) | (-1.214) | (-0.724) | (0.362) | **(-2.316)** | (-0.220) | (0.654) | (-0.359) |
| FDI/GDP | 0.0051* | 0.004*** | 0.024 | 0.031** | 0.003** | 0.010*** | 0.003*** | 0.002*** |
|  | (1.812) | (4.267) | (0.341) | (2.990) | (2.201) | (5.498) | (4.380) | (4.138) |
| INV/GDP | 0.1018* | -0.141 | 0.203* | 0.028** | 0.131 | 0.054** | 0.110* | -0.005 |
|  | (1.7054) | (-1.504) | (1.698) | (-2.505) | (0.524) | (2.789) | (1.758) | (-0.203) |
| OPEN | 0.0962** | -0.009 | 0.293*** | -0.019 | 0.122 | 0.115* | 0.125* | 0.116* |
|  | (2.506) | (-0.185) | (4.053) | (-0.604) | (1.680) | (1.813) | (1.722) | (1.806) |
| Credits/GDP | 0.1168 | 0.041 | 0.167 | -0.022 | -0.046 | 0.136 | 0.076 | 0.113 |
|  | (1.005) | (0.588) | (1.225) | (-0.680) | (-0.511) | (0.404) | (0.534) | (0.814) |
| REER | -0.2273* | 0.244 | 0.132 | 0.098 | 0.241 | 0.004 | -0.369* | -0.252** |
|  | (-1.794) | (0.962) | (0.320) | (0.472) | (1.084) | (0.012) | (-1.903) | (2.593) |
| Cragg-Donald F-statistic | 36.29 | 32.17 | 34.49 | 41.05 | 36.78 | 24.21 | 30.16 | 29.48 |
| | **Whole period (1990:Q1-2015:Q3)** | | | | | | | |
| C | -3.2569 | -1.796 | 1.144 | -27.759 | -17.953 | -48.916 | -13.690 | 15.511 |
|  | (-1.389) | (-0.756) | (0.157) | (-0.819) | (-0.753) | (-0.370) | (-0.462) | (0.279) |
| REM/GDP | 0.0192* | 0.027 | 0.456 | 0.273 | 0.223 | **0.050\*** | **0.035\*\*** | **0.127\*\*\*** |
|  | (1.832) | (0.105) | (0.266) | (0.568) | (0.343) | **(1.739)** | **(2.525)** | **(3.433)** |
| FDI/GDP | -0.0772** | -0.067* | -0.569 | -0.215 | -0.114* | -0.09*** | 0.011*** | -0.098 |
|  | (-2.694) | (-1.789) | (-1.125) | (-1.176) | (-1.897) | (-3.742) | (4.292) | (-1.954) |
| INV/GDP | 0.0298* | 0.010 | 1.129 | 0.082* | 0.076** | 0.023 | 0.067 | 0.061* |
|  | (1.794) | (0.835) | (1.186) | (1.911) | (2.589) | (1.414) | (0.925) | (1.911) |
| OPEN | 0.0892** | 0.097** | 0.102* | 0.089*** | 0.115** | 0.114*** | 0.098** | 0.038* |
|  | (2.567) | (2.546) | (1.956) | (3.972) | (2.756) | (2.913) | (2.765) | (1.816) |
| Credits/GDP | 0.0342* | 0.021* | 0.010 | 0.031*** | 0.045** | 0.026*** | 0.047 | 0.035*** |
|  | (1.801) | (1.713) | (1.365) | (3.009) | (2.879) | (5.139) | (1.251) | (3.818) |
| REER | -0.1945*** | -0.081** | -0.123 | -0.045* | -0.606 | -0.168 | -0.362 | -0.285 |
|  | (-3.189) | (-2.695) | (-0.657) | (-1.923) | (-0.865) | (-0.924) | (-1.415) | (-0.717) |
| Cragg-Donald F-statistic | 34.89 | 36.21 | 31.67 | 31.72 | 34.07 | 25.28 | 32.18 | 30.89 |

Notes: ***, ** and * imply significance at the 1%, 5% and 10%, respectively. 10% and 15% critical value of Stock–Yogo weak idetification test are 17.02 and 13.85, respectively; The null hypothesis of weak instruments or Cragg–Donald F-statistic test can be rejected when the associated F-statistic values appear stronger than the critical values by thresholds provided by Stock and Yogo (2005).

---

[11] Instead of using time domain analysis allowing to analyze the relationship between remittances and macroeconomic variables throughout the entire period, the correlation analysis-based EMD permits to see how behave the investigated linkage across various time-scales.

**Table 9. Regression of investment on remittances (Control for endogeneity)**

| | Time domain | IMF1 | IMF2 | IMF3 | IMF4 | IMF5 | IMF6 | IMF7 |
|---|---|---|---|---|---|---|---|---|
| | **Restricted period (1990:Q1-2010:Q4)** | | | | | | | |
| C | -6.542** | -9.321* | -8.324 | -6.7*** | -6.739*** | -7.459* | -7.212** | -6.542*** |
| | (-2.345) | (-1.894) | (-1.234) | (-7.234) | (-3.513) | (-1.867) | (-2.852) | (-4.510) |
| REM/GDP | -0.039** | **0.062*** | **-0.134*** | -0.074* | 0.146 | 0.102 | 0.056 | 0.105 |
| | (-2.632) | **(1.796)** | **(-3.865)** | (-1.891) | (0.975) | (1.136) | (1.189) | (1.128) |
| FDI/GDP | -0.009** | -0.061 | 0.073 | -0.303 | -1.006 | -0.07*** | -0.06*** | -0.083** |
| | (-2.684) | (-1.238) | (1.234) | (-1.278) | (-1.135) | (-3.291) | (-4.011) | (-2.612) |
| gGDP | 0.369 | 0.819 | -0.256 | 0.274 | 0.139 | 0.124 | 0.436 | 0.185 |
| | (1.045) | (0.227) | (-1.616) | (0.812) | (1.000) | (1.514) | (1.048) | (1.313) |
| OPEN | 0.107* | 0.079 | 0.081* | 0.164 | 0.210 | 0.131 | 0.146** | 0.137* |
| | (1.863) | (1.426) | (1.891) | (1.424) | (1.358) | (1.976) | (2.525) | (1.924) |
| Credits/GDP | 0.035* | -0.151 | -0.167 | 0.129 | 0.472 | 0.031* | 0.026*** | 0.022*** |
| | (1.942) | (-1.303) | (-1.411) | (1.101) | (0.869) | (1.756) | (4.158) | (3.194) |
| CPI | -0.168* | -0.135* | -0.678 | -0.105 | -0.367 | -0.225* | -0.171* | -0.215** |
| | (-1.875) | (-1.912) | (-1.193) | (-1.247) | (-1.235) | (-1.834) | (-1.912) | (-2.472) |
| RIR | -0.094** | -0.178* | -0.092 | -0.076* | -0.023* | -0.129 | -0.182 | -0.045** |
| | (-2.352) | (-1.796) | (-1.414) | (-1.543) | (-1.703) | (-0.738) | (-0.503) | (-1.615) |
| Cragg-Donald F-statistic | 26.79 | 25.67 | 26.71 | 26.72 | 28.32 | 28.01 | 28.00 | 29.12 |
| | **Whole period (1990:Q1-2015:Q3)** | | | | | | | |
| C | -4.503** | -3.792*** | -4.123** | -4.58*** | -4.096** | -4.100* | -4.196** | -4.811** |
| | (-2.689) | (-4.525) | (-2.525) | (-4.515) | (-2.323) | (-1.891) | (-2.613) | (-2.356) |
| REM/GDP | -0.129 | -0.145* | **-0.136*** | 0.063 | 0.045 | 0.021 | 0.356 | 0.142 |
| | (-1.639) | (-1.909) | **(-1.811)** | (1.286) | (0.678) | (0.616) | (1.325) | (0.796) |
| FDI/GDP | -0.061** | -0.156 | 0.368 | -0.245 | -0.358 | -0.062 | -0.08*** | -0.056* |
| | (-2.342) | (-1.103) | (0.511) | (-1.567) | (-1.034) | (-1.245) | (-3.629) | (-1.869) |
| gGDP | 0.067*** | -0.621 | 0.421 | 0.094** | 0.156 | 0.092* | 0.088* | 0.122 |
| | (3.109) | (-0.855) | (1.236) | (2.678) | (1.245) | (1.956) | (1.875) | (1.074) |
| OPEN | 0.051* | 0.039* | 0.098*** | 0.164 | 0.167 | 0.234 | 0.065* | 0.113 |
| | (1.768) | (1.892) | (3.819) | (1.424) | (1.023) | (1.126) | (2.100) | (1.045) |
| Credits/GDP | 0.038* | -0.009 | 0.076* | 0.129 | 0.067 | 0.028*** | 0.138 | 0.156 |
| | (1.910) | (-1.134) | (1.810) | (1.101) | (1.008) | (3.896) | (1.249) | (0.689) |
| CPI | -0.083** | -0.095*** | -0.096** | -0.105 | 0.135 | -0.131 | -0.138 | -0.034* |
| | (-2.819) | (-3.621) | (-2.553) | (-1.247) | (0.921) | (-1.424) | (-1.256) | (-1.826) |
| RIR | -0.045** | -0.038* | -0.515 | -0.051* | -0.312 | -0.085* | 0.096 | -0.005 |
| | (-2.378) | (-1.864) | (-1.123) | (-1.747) | (-0.767) | (-1.698) | (-1.002) | (-0.912) |
| Cragg-Donald F-statistic | 24.56 | 23.15 | 29.07 | 26.15 | 28.14 | 21.87 | 22.13 | 25.67 |

Notes: ***, ** and * imply significance at the 1%, 5% and 10%, respectively. 10% and 15% critical value of Stock–Yogo weak idetification test are 17.02 and 13.85, respectively; The null hypothesis of weak instruments or Cragg–Donald F-statistic test can be rejected when the associated F-statistic values appear stronger than the critical values by thresholds provided by Stock and Yogo (2005).

**Table 10. Regression of Consumption on remittances (Control for endogeneity)**

|  | Time domain | IMF1 | IMF2 | IMF3 | IMF4 | IMF5 | IMF6 | IMF7 |
|---|---|---|---|---|---|---|---|---|
|  | | **Restricted period (1990:Q1-2010:Q4)** | | | | | | |
| C | 3.109*** | 2.891*** | 1.783* | 2.672** | 1.924*** | 2.345*** | 1.356** | 1.642*** |
|  | (3.571) | (3.912) | (1.881) | (2.936) | (3.876) | (4.516) | (1.972) | (3.514) |
| REM/GDP | 0.067** | **0.103*** | **0.095*** | 0.114 | 0.167 | 0.279 | 0.313 | 0.129 |
|  | (2.871) | **(1.834)** | **(2.717)** | (1.639) | (1.056) | (0.832) | (1.042) | (1.361) |
| Credits/GDP | 0.068** | 0.092 | 0.126*** | 0.119*** | 0.549 | 0.212 | 0.083* | 0.215 |
|  | (2.425) | (1.414) | (4.267) | (3.125) | (1.309) | (1.192) | (1.914) | (1.318) |
| gGDP | 0.029*** | 0.023** | 0.011* | 0.014* | 0.097* | 0.017* | 0.069 | 0.108* |
|  | (3.814) | (2.511) | (1.912) | (1.822) | (2.064) | (1.695) | (1.254) | (1.935) |
| CPI | -0.181*** | -0.214** | -0.156** | -0.194*** | -0.167** | -0.212*** | -0.162* | -0.171** |
|  | (-3.619) | (-2.356) | (-3.004) | (-4.361) | (-1.982) | (-4.918) | (-1.724) | (-2.526) |
| RIR | -0.037** | -0.023 | -0.011* | -0.049 | 0.014 | -0.044* | -0.052* | -0.021** |
|  | (-2.618) | (-1.414) | (-1.749) | (-1.020) | (1.187) | (-1.781) | (-1.912) | (-2.814) |
| Cragg-Donald F-statistic | 22.35 | 19.87 | 21.42 | 20.98 | 21.15 | 22.37 | 23.14 | 21.68 |
|  | | **Whole period (1990:Q1-2015:Q3)** | | | | | | |
| C | 3.892* | 4.056 | 3.246 | 4.156*** | 4.092** | 4.156*** | 3.565** | 3.916*** |
|  | (1.716) | (1.127) | (1.214) | (3.672) | (2.627) | (3.492) | (2.482) | (3.227) |
| REM/GDP | 0.129* | 0.062 | 0.131 | 0.114 | 0.110 | 0.098 | **0.135*** | **0.126**** |
|  | (1.876) | (1.378) | (0.907) | (1.316) | (1.458) | (1.945) | **(2.691)** | **(3.711)** |
| Credits/GDP | 0.056*** | -0.01*** | 0.096 | 0.054 | 0.076 | 0.049 | 0.069*** | 0.088* |
|  | (3.809) | (-4.213) | (1.154) | (1.319) | (1.020) | (1.286) | (4.712) | (1.972) |
| gGDP | 0.034** | 0.021* | 0.023** | 0.076 | 0.100 | 0.094 | 0.045* | 0.094** |
|  | (2.573) | (1.699) | (2.167) | (1.434) | (0.928) | (1.518) | (1.758) | (2.076) |
| CPI | -0.186** | -0.28*** | -0.197** | -0.234*** | -0.256** | -0.267* | -0.245** | -0.189* |
|  | (-2.714) | (-4.312) | (-2.652) | (-3.861) | (-2.489) | (-1.993) | (-2.417) | (-1.762) |
| RIR | -0.066*** | -0.072* | -0.096** | -0.110** | -0.08*** | -0.072** | -0.105* | 0.141 |
|  | (-3.298) | (-1.914) | (-2.527) | (-2.314) | (-4.112) | (-2.334) | (-1.921) | (-1.196) |
| Cragg-Donald F-statistic | 23.18 | 21.04 | 20.18 | 19.76 | 18.34 | 17.26 | 20.13 | 19.82 |

Notes: ***, ** and * imply significance at the 1%, 5% and 10%, respectively. 10% and 15 % critical value of Stock–Yogo weak idetification test are 17.02 and 13.85, respectively; The null hypothesis of weak instruments or Cragg–Donald F-statistic test can be rejected when the associated F-statistic values appear stronger than the critical values by thresholds provided by Stock and Yogo (2005).

### 6.2. The frequency domain causality results

A further step for robustness check consists of employing frequency domain causality test [12] to test whether there is a causal relationship between remittances and the focal macroeconomic variables from one frequency to another. The figure contains the test statistics with their 5 percent critical values for the frequency bands involved (solid line) over the interval [0, π]. The frequency ($\omega$) on the horizontal axis can be translated into a cycle or periodicity of $T$ weeks by $T = (2\pi/\omega)$ where $T$ is the period. Figure 7.1 describes the evolution of the causal relationship between growth and remittances depending to frequency transformations. Before the Arab Spring, we support a medium-run unidirectional causality from remittances to growth, especially when $\omega \in [1.30\pi - 2.60\pi]$ corresponding to a cycle length between 2.4 and 4.5 quarters. However, a long-run causal relation running from remittances to growth happened when focusing on the whole period (before and after the Arab Spring), in particular when $\omega \in [0.01\pi - 0.98\pi]$, corresponding to a cycle superior to 6.4 quarters. The reverse link is not validated at any frequency and at any estimation period.

---

[12] For details about the procedure of this technique, you can refer to Overview A.1 (Appendices).

For the remittances impact on investment, a slight change was marked by moving from the restricted to the whole period (Figure 7.2). Prior to Arab Spring event, remittance inflows Granger-caused domestic investment in high frequencies (when $\omega \in [2.81\pi - 3.03\pi]$, in particular for a cycle less than 2.2 quarters). This relationship remained driven by quickly fluctuating components for the whole period. Nevertheless, the remittances' impact on INV/GDP was stronger for the second period as the cycle expands to 2.7 quarters ($\omega \in [2.27\pi - 3.03\pi]$).

As the cycle remittances-growth (Figure 6.1), a causal link running from remittances to consumption (Figure 7.3) was supported in the short-run for the restricted period (when $\omega \in [2.70\pi - 3.03\pi]$), corresponding to a cycle length inferior to 2.3 quarters) and in the long-run for the whole period (when $\omega \in [0.01\pi - 0.54\pi]$) corresponding to a cycle above 11.6 quarters).

Overall, the frequency domain causality findings seem consistent with the correlation analysis-based EMD. Specifically, the consideration of the Arab spring period in our estimates led to sharp changes in the cycles remittances-growth and remittances-consumption; the cycles which were valid in the short term for the restricted period, became explained by long-term oscillations for the whole period. This confirms the consistency of these two cycles. The cycle remittances-investment and remittances-real effective exchange rate changed too but moderately. However, the linkage between REM/GDP and INV/GDP remained driven by short-term factors for the two periods.

**Figure 7. The frequency domain causality between remittances and macroeconomic variables**

| Restricted period (1990:Q1-2010:Q4) | Whole period (1990:Q1-2015:Q3) |

### 7.1. REM/GDP and gGDP

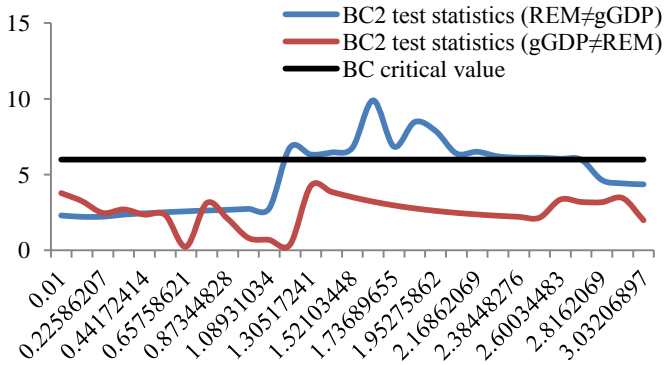
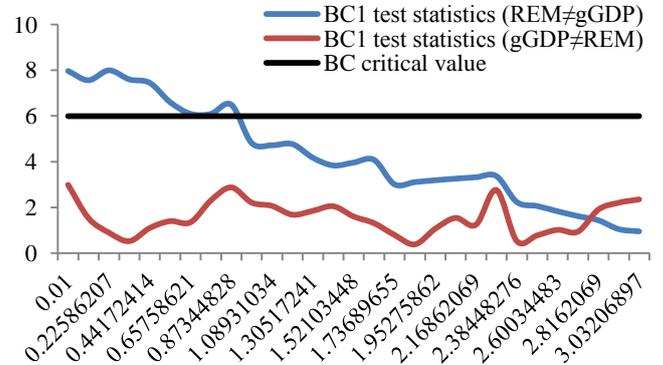

### 7.2. REM/GDP and INV/GDP

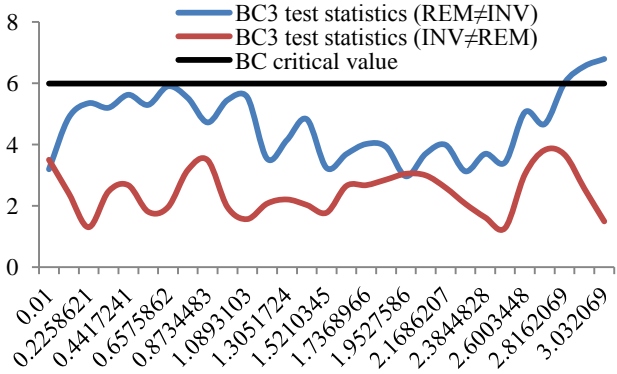
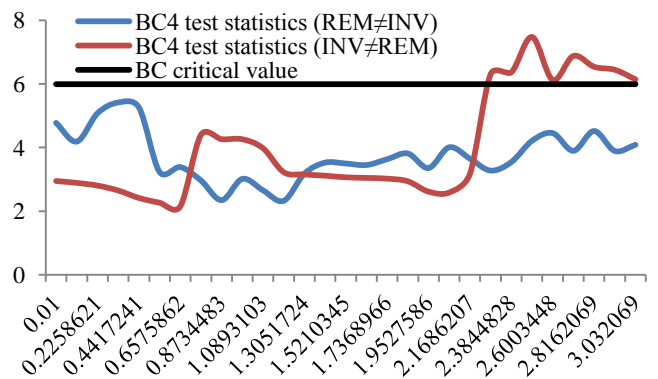

### 7.3. REM/GDP and CONS/GDP

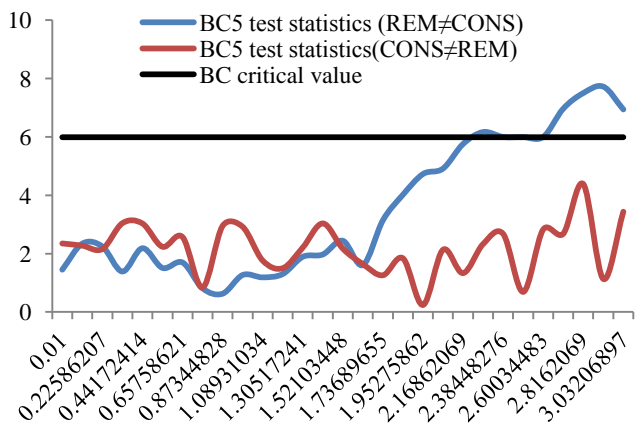
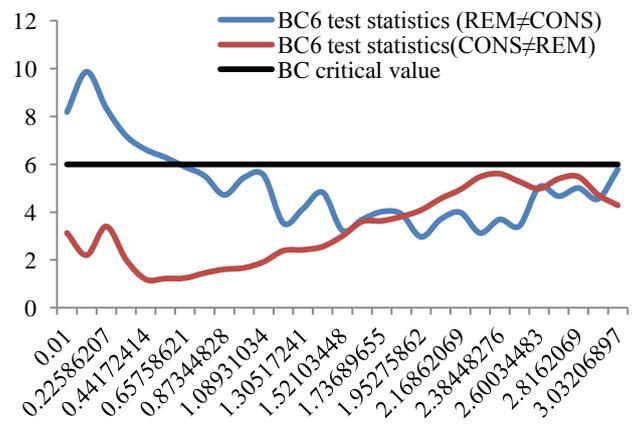

# 7. Conclusion and policy implications

Before the downfall of Ben Ali's regime, Tunisia has succeeded to have a prosperous economy but the country is destabilized by the Arab Spring that came to underscore the weakness of the pillars of its economy which had not withstood this shock. The tourism collapsed, FDI dried up, the foreign trade did not resist, and the dinar depreciated. Unusually, remittances survived and even rose, highlighting their countercyclical behavior. In light of this observation, this study attempts to determine the channels through which these financial inflows can help to boost economic growth in a country that saw extreme social and political turmoil. This article uses newly econometric techniques which contain several novel features that set this study apart from the literature on the issue. We use a multi-scale analysis based on Empirical Mode Decomposition. This method aims at disentangling each variable into different scaling components and at each scale estimating the correlation between the variables under study. These methods allow us to extract intrinsic features inherent to the time series. This is expected to yield more accurate and minute scrutiny which would estimate "complex" relationship between remittances and macroeconomic variables, i.e., economic growth, domestic investment and consumption in an unstable context.

Because we have not enough observations to make estimation for the post-Arab spring period, we thought to consider (1) a restricted period: prior to the aftermath of Arab Spring and (2) a whole or extended period: before and after the onset of Arab Spring event. Despite this limitation, three relevant outcomes are drawn. First, although the remittances' effect on growth is negative and dominantly determined by short-term inner factors in the restricted period, it becomes positive and driven by long-term factors in the extended period. Second, while in restricted period, the remittances' impact on investment is likely to be variant (negative in some scales and positive in others) and explained by short-term inner features, in the extended period, this effect becomes negative, weak and driven by short- and medium-term factors. Third, migrants' remittances have a positive and significant effect on consumption in the two periods. But, this effect is held in the short-run in the restricted period and in the long-run in the whole period.

These findings suggest that it is unnecessary to oppose the two transmission channels (consumption vs. investment) through which remittances can significantly affect Tunisia's growth. In particular, remittances are found to be driven by the need to support migrant worker's families rather than by investment considerations. This suggests the importance of remittances as coping mechanism against shocks, without typically turning the recipients into investors, stimulating entrepreneurial activities, rising formal sector employment, and generating multiplier effect [13]. Even if they not be used "productively", a positive and long-run impact on growth appears robust. In time of crisis, remittances will help families to heal and to continue sending their children to school. This type of behaviour may be potentially reflecting, in certain circumstances, a preferable investment for the families. However, within the context of political and social unrest investors' disquiets over the economic prospects of this country exacerbate, harming the investment climate. What is noticeable these last weeks, however, is that Tunisians are witnessing a sharp devaluation of dinar. As a result, the export-competing companies would be harmfully influenced by the real exchange rate overvaluation and the related potential loss of international competitiveness. The adverse effects of the loss in external competitiveness can be mitigated by stimulating the internal competitiveness. The Tunisian authorities should open different economic sectors to competition, develop a fair administrative business environment and undertake proactive reforms, tax benefits, organization, governance mechanisms and other regulations to strengthen the involvement of Tunisians abroad in national development process. Through a new legal and institutional framework for investment [14] (law n°71 of September 30, 2016), Tunisia aims to overcome the long-winded economic difficulties, to change the current economic model and to adopt a new economic model based on efficiency and productivity via the encouragement of investment in innovative sectors and sectors with higher value- added and the enhancement of export capacity and technological content of the Tunisian economy. This would help to boost the competitiveness of the national economy and mitigate the low employment rate and the country's regional disparities. Tunisia is at a turning point today, facing multiple challenges but also aspiring to potential opportunities. The new law is expected to stimulate investment environment and market opportunities for businesses in Tunisia.

Last but not least, on the basis of this article' findings, we cannot affirm that the remittances flows are able to fully cushion the uncertainty surrounding the current Tunisia's situation. Certainly they have increased remarkably, affecting directly both the balance of payment and the wellbeing of families who receive them, but this situation is exogenous and their total impact will depend on policy measures taken to encourage them. Once political stability is achieved, a special attention is needed for channeling remittance inflows towards productive investments. This requires learning more about the range of barriers to using them for investment and the effective institutions that can effectively guide recipients of remittances make the most of the remittances they receive.

---

[13] This can happen but in infrequent cases. Papers that focused on the impact of remittances on investment seem very scarce. Generally, migrants come to invest in their country of origin on condition that they monitor themselves their investment. For a summary of these studies, you can refer to Bouoiyour et al. (2016).
[14] For more details about the new law, you can visit this link: http://www.ilboursa.com/marches/tunisie-les-principales-caracteristiques-du-nouveau-cadre-juridique-de-l-investissement_11291

**References**

Aggarwal, R., Demirgüç-Kunt, A and Pería, M.S.M (2006). Do workers' remittances promote financial development? World Bank Policy Research Working Paper 3957.

Amuedo-Dorantes, C and Pozo, S (2004). Workers' remittances and the real exchange rate: A paradox of gifts. World Development, 32(8), 1407-1417.

Bouoiyour J. (2006). Migration, Diaspora et développement humain, in « Le Maroc possible, une offre de débat pour une ambition collective » Rapport du cinquantenaire, 2006. Royaume du Maroc.
http://www.rdh50.ma/fr/pdf/contributions/GT3-8.PDF

Bouoiyour, J, Miftah A.and Muller C. (2016). Remittances and Living Standards in Morocco. Working paper, CATT, University of Pau.Bouoiyour, J. and Selmi, R. (2016 ). Dutch Disease, Remittances and Arab Spring in Tunisia, Working paper, CATT, University of Pau.

Breitung, J and Candelon, B. (2006). Testing for short and long-run causality: a frequency domain approach. Journal of Econometrics 132, 363-378.

Camarero, M. and Tamarit, C. (2002). Oil prices and Spanish competitiveness: A cointegrated panel analysis. Journal of Policy Modeling 24(6), 591-605.

Carrera, J. E. and Restout, R. (2008). Long Run Determinants of Real Exchange Rates in Latin America. Working paper GATE 08-11.

Central Bank of Tunisia (2014). La balance des paiements et la position extérieure globale de la Tunisie 2013.

Chami, R, Fullenkamp, C, and Jahjah, S. (2005). Are immigrant remittance flows a source of capital for development? IMF Staff Papers, 52(1), 55-81.

Chnaina, K and Makhlouf, F. (2015). Impact des Transferts de Fonds sur le Taux de Change Réel Effectif en Tunisie. African Development Review. DOI: 10.1111/1467-8268.12130

Barajas, A, Chami, R, Fullenkamp, C, Gapen, M and Montiel, M. (2009). Do workers' remittances promote economic growth? IMF Working Paper 09/153, Washington, D.C. International Monetary Fund.

Drazin, P. G. (1992). Nonlinear systems. Cambridge University Press.

Geweke, J. (1982). Measurement of linear dependence and feedback between multiple time series. Journal of American Statistical Association 77, 304-324.

Granger, C.W.J. (1969), Investigation causal relations by econometric models and cross-spectral methods. Econometrica 37, 424-438.

Gubert , F., Lassourd, T. and Mesplé-Somps, S. (2010). Do remittances affect poverty and inequality? Evidence from Mali. Dauphine Working Paper.

Giuliano, P and Ruiz-Arranz, M. (2009). Remittances, financial development, and growth. Journal ofDevelopment Economics 90, 144-152.

Hassan, G. Mainul., Shakur, S. and Bhuyan, M. (2012). Nonlinear growth effect of remittances in recipient countries: an econometric analysis of remittances-growth nexus in Bangladesh. Annual NZAE Conference (pp. 1-35). New Zealand Association of Economists.

Huang, N.E, Z. Shen, S. R. Long, M. C. Wu, H. H. Shih, Q. Zheng, N.-C. Yen, C. C. Tung, and H. H. Liu. (1998). The Empirical Mode Decomposition and Hilbert Spectrum for Nonlinear and Nonstationary Time Series Analysis. Proceedings of the Royal Society London A., 454:903–995.Huang, N.E.,Wu, M.L., Long, S.R., Shen, S.S.P.,Qu,W.D.,Gloersen, P., Fan, K.L., (2003). A confidence limit for the empirical mode decomposition and the Hilbert spectral analysis. Proceedings of the Royal Society of London. A 459, pp. 2317–2345.

Huang, N. E. and N. O. Attoh-Okine, (2005). Hilbert-Huang Transform in Engineering. CRC Press.IOM (2011). Quels liens les tunisiens résidant en Europe gardent-ils avec le pays d'origine ? In the project TIDO "Tunisian migrants Involved in Development of the country of Origin".

Jouini, J. (2015). Economic growth and remittances in Tunisia: bi-directional causal links. Journal of Policy Modeling, http://dx.doi.org/10.1016/j.jpolmod.2015.01.015.

Kifle, T. (2007). Do Remittances Encourage Investment in Education? Evidence fromEritrea. Journal of African studies, 4(1).Kouni, M., (2016). Remittances and Growth in Tunisia: A Dynamic Panel Analysis from a Sectoral Database. Journal of Emerging Trends in Economics and Management Sciences, 7(5), 342-351.

Lin, L., and J. Hongbing, J. (2009). Signal feature extraction based on an improved EMD method", Measurement, 42 (5), 796-803.

Litterman, R.B. and Weiss, L. (1983). Money, real interest rates, and output: a reinterpretation of postwar U.S. data. NBER working paper series n°1077.

Lueth E and Ruiz-Arranz, M. (2007). Are Workers' Remittances a Hedge Against Macroeconomic Shocks? The Case of Sri Lanka. IMF, WP/ 07/22.

Mansuri, G. (2006). Migration, school attainment, and child labor: evidence from rural Pakistan. Policy Research Working Paper Series 3945, World Bank.

McCormick, B and Wahba, J. (2003). Overseas work experience, savings and entrepreneurship amongst return migrants to LDCs. Journal of African economies, 12 (4), 500-532.

Mesnard, A. (2004). Temporary migration and capital market imperfections. Oxford Economic Papers, 56(2), 242-262.Ng, S. and Perron, P. (2001). Lag Length Selection and the Construction of Unit Root Tests with Good Size and Power. Econometrica 69(6), 1519-1554.

Fayissa, B and Nsiah, C. (2008). The impact of remittances on economic growth and development in Africa. Working Paper Series, Middle Tennessee State University.

Fayissa, B and Nsiah, C. (2010). Can remittances spur economic growth and development? Evidence from Latin American Countries. Department of Economics and Finance Working Paper Series, Middle Tennessee State University.

Glytsos, N.P. (2002). Dynamic effects of migrant remittances on growth: An Econometric Model with an application to Mediterranean Countries. Centre Of Planning And Economic Research No 74.

Özden, C and Schiff, M. (2006). Overview. In C. Özden & M. Schiff (Eds.), International migration, remittances & the brain drain (1-18). Washington, DC: World Bank and Palgrave Macmillan.

OECD (2013). World Migration in Figures. A joint contribution by UN-DESA and the OECD to the United Nations High-Level Dialogue on Migration and Development, 3-4 October.

Pollack, K.M. (2011). Understanding the Arab awakening. http://www.brookings.edu/~/media/press/books/2011/thearabawakening/thearabawakening_chapter.pdf

Rao, B. B and Hassan, G M (2012). Are Direct and Indirect Growth Effects of Remittances Significant? The World Economy. Vol 35(3), 351 – 372.

Ruiz, I, Shukralla, E and Vargas-Silva C (2009). Remittances, Institutions and Growth: A Semiparametric Study. International Economic Journal, 23(1), 111-119.

Sims, C.A. (1980). Macroeconomics and Reality. Econometrica, 48, 1-48.

Sun, E. W. and T. Meinl (2012). A new wavelet-based denoising algorithm for highfrequency financial data mining. European Journal of Operational Research 217 (3), 589–599.

Valero-Gil, J (2008). Remittances and the household's expenditures on health. MPRA Paper 9572, University Library of Munich, Germany.

Woodruff, C and Zenteno, R. (2004). Remittances and microenterprises in Mexico. Graduate School of International Relations and Pacific Studies Working Paper.

# Appendices

**Figure A.1. The IMFs involved of gGDP, REM/GDP, INV/GDP and CONS/GDP**

| **Restricted period (2000:Q1-2010:Q4)** | **Whole period (2000:Q1-2015:Q3)** |
|---|---|

gGDP

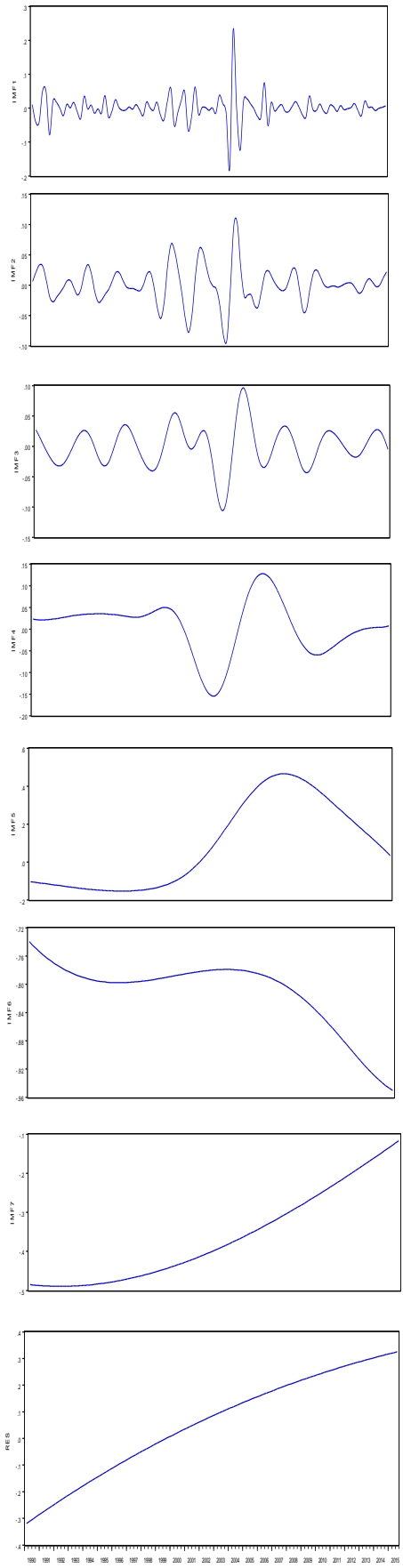
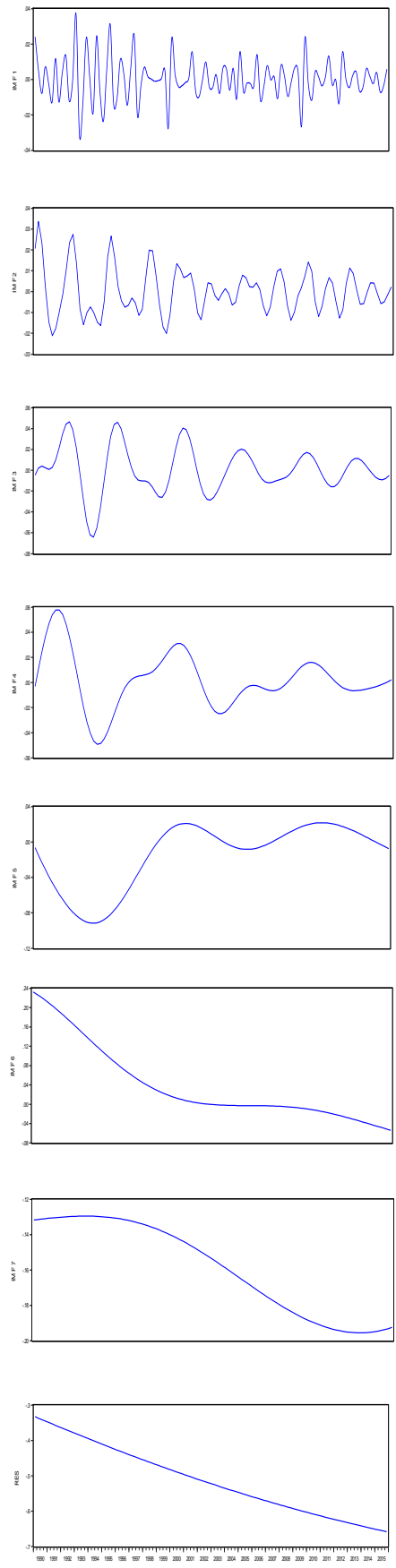

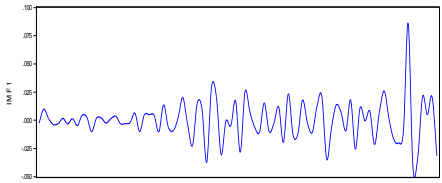
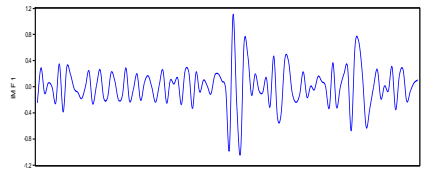
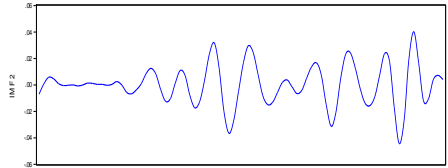
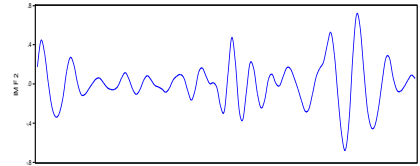
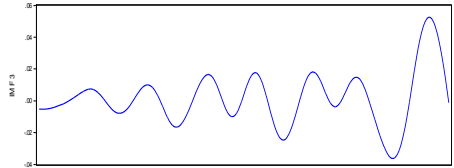
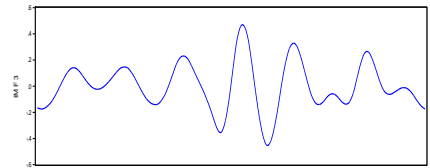
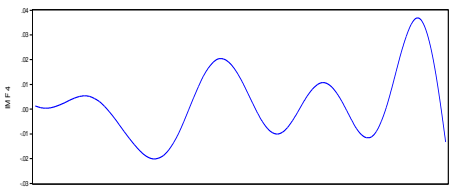
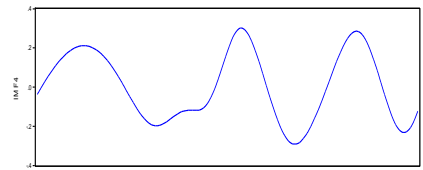
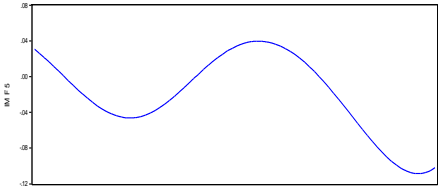
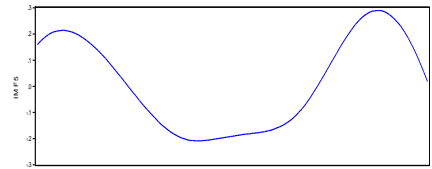
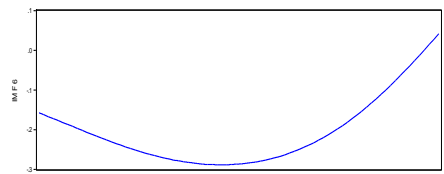
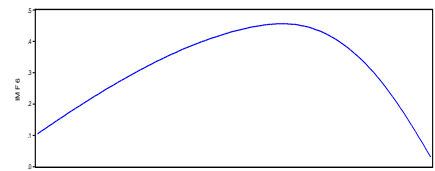
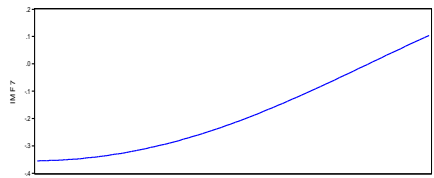
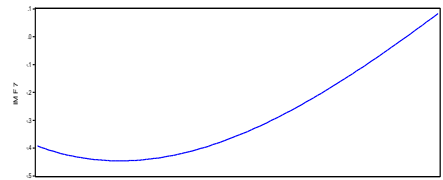
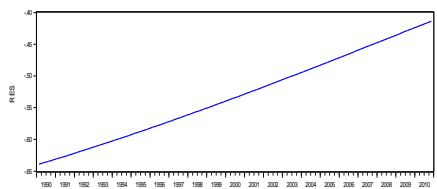
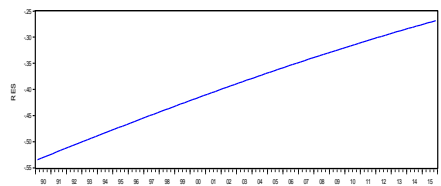

CONS/GDP

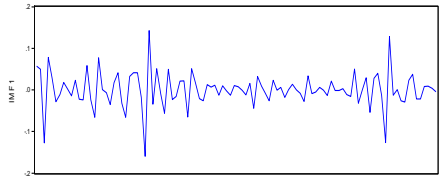
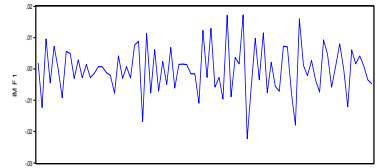
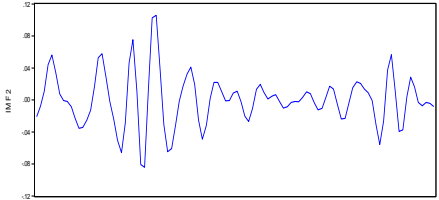
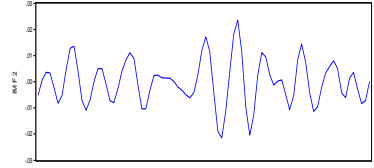
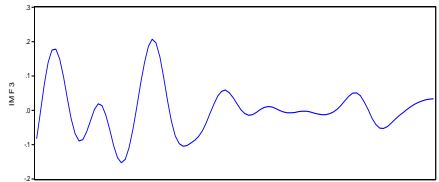
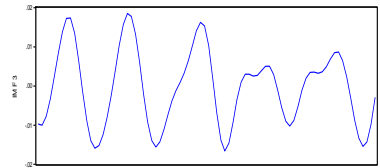
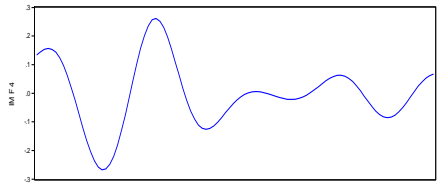
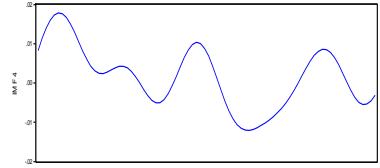
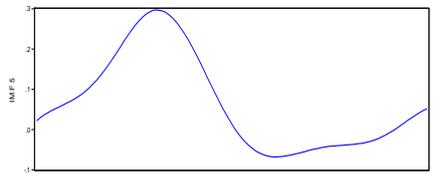
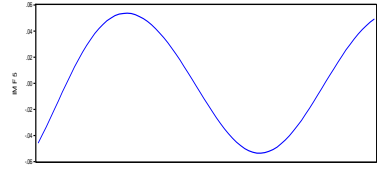
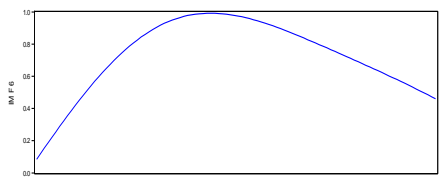
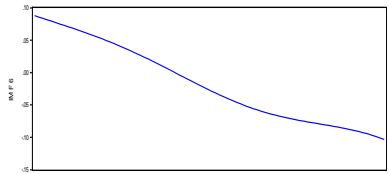
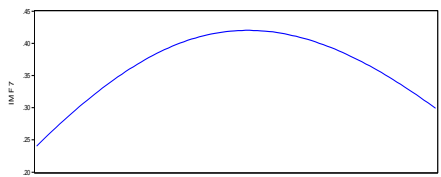
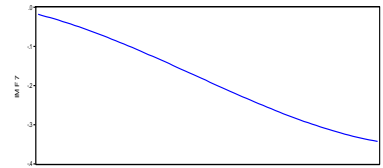
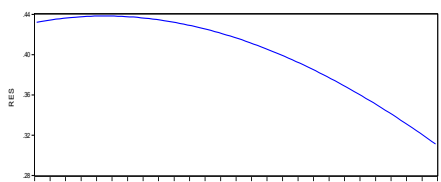
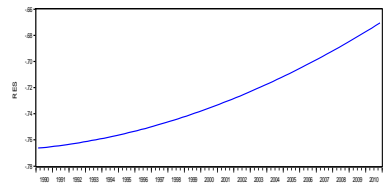

**Overview A.1. A frequency domain causality test**

The present paper made an attempt to assess the causal linkage between remittances and macroeconomic variables through a recently developed signal approach of Breitung and Calderon (2006). Use of this approach allows disentangling the Granger causality in the frequency domain, and then identify if the predictive power is concentrated at the quickly fluctuating components (high frequency) or at the slowly fluctuating components (low frequency). This distinction is very important in studying causality. Although conceptually interesting, the standard Granger causality test does not permit discerning the variant characteristics of the signals involved and which normally play a significant role on the underlying series; hence the usefulness of the decomposition of data variables into various frequencies that may help policy makers in the formulation of the adequate decisions. Precisely, the covariance of these variables is disentangled into various spectral components. The aim is that a stationary process can be depicted as a weighted sum of sinusoidal components with a certain frequency, enabling to evaluate the underlying cyclical properties of the times series studied and the linkage between them.

To review the testing procedure, let us suppose that a two-dimentional time series vector $[x_t, y_t]$ is generated by the following stationary VAR(p) model:

$$\begin{pmatrix} x_t \\ y_t \end{pmatrix} = \begin{pmatrix} \phi_{11}(L) & \phi_{12}(L) \\ \phi_{21}(L) & \phi_{22}(L) \end{pmatrix} \begin{pmatrix} x_{t-1} \\ y_{t-1} \end{pmatrix} + \begin{pmatrix} \mu_t \\ \upsilon_t \end{pmatrix} = \begin{pmatrix} \psi_{11}(L) & \psi_{12}(L) \\ \psi_{21}(L) & \psi_{22}(L) \end{pmatrix} \begin{pmatrix} \varepsilon_t \\ \eta_t \end{pmatrix}, t=1,\ldots, T \quad (A.1)$$

where $\phi_{ij}(L) = \phi_{ij,1} L^0 + \ldots + \phi_{ij,p} L^{p-1}$ for $i,j = 1,2$ and $[\mu_t, \upsilon_t] \sim iid(0, \sum)$. Note that is positive definite and let G be the lower triangular matrix of the Cholesky $G'G = \sum^{-1}$; $[\varepsilon_t, \eta_t]'$ is defined as $G[\mu_t, \upsilon_t]'$ and $\psi_{ij}(L)$ for $i,j = 1,2$ are defined accordingly.

Then, the population spectrum of $x$, denoted by $f_x(\omega)$, can be derived from the previous matrix and expressed as follows:

$$f_x(\omega) = \frac{1}{2\pi} \left\{ \left| \psi_{11}(e^{-i\omega}) \right|^2 + \left| \psi_{12}(e^{-i\omega}) \right|^2 \right\} \quad (A.2)$$

The main goal of this technique is to test whether $x_t$ Granger cause $y_t$, at a given frequency λ, even if we control for $Z_t$ ( additional control variables). Geweke (1982) developed a measure of causality denoted as:

$$M_{x \to y/Z}(w) = \log\left[1 + \frac{|\psi_{12}(e^{-iw})|^2}{|\psi_{11}(e^{-iw})|^2}\right] \quad (A.3)$$

As $|\psi_{12}(e^{-iw})|^2$ is a complex function of the VAR parameters, Breitung and Candelon (2006) and in order to resolve this drawback, argued that the hypothesis $M_{x \to y/Z}(\omega) = 0$ correspond to a linear restriction on the VAR coefficients.

$$H_0 : R(\omega)\phi(L) = 0 \quad (A.4)$$

where $R(\omega) = \begin{bmatrix} \cos(\omega)\cos(2\omega)...\cos(p\omega) \\ \sin(\omega)\sin(2\omega)...\sin(p\omega) \end{bmatrix}$

The significance of the causal relationship can be tested by a standard F-test or by comparing the causality measure for $\omega \in [0, \pi]$ with the critical value of a $\chi^2$ distribution with 2 degrees of freedom, which is 5.99.